\documentclass[conference]{IEEEtran}
\IEEEoverridecommandlockouts
\usepackage{amsmath,amssymb,amsfonts}
\usepackage{algorithmic}
\usepackage{graphicx}
\usepackage{textcomp}
\usepackage{xcolor}
\usepackage{tikz}
\def\BibTeX{{\rm B\kern-.05em{\sc i\kern-.025em b}\kern-.08em
    T\kern-.1667em\lower.7ex\hbox{E}\kern-.125emX}}

\usepackage[backend=biber,  
style=numeric,      
maxbibnames=99,      
minalphanames=3,      
backref=false        
]{biblatex}
\bibliography{bibfile.bib}

\begin{document}

\title{Ksurf: Attention Kalman Filter and Principal Component Analysis for Prediction under Highly Variable Cloud Workloads
}

\author{\IEEEauthorblockN{1\textsuperscript{st} Michael Dang'ana}
\IEEEauthorblockA{\textit{Electrical \& Computer Engineering} \\
\textit{University of Toronto}\\
Toronto, Canada \\
michael.dangana@mail.utoronto.ca}
\and
\IEEEauthorblockN{2\textsuperscript{nd} Arno Jacobsen}
\IEEEauthorblockA{\textit{Electrical \& Computer Engineering} \\
\textit{University of Toronto}\\
Toronto, Canada \\
jacobsen@eecg.toronto.edu}
}

\maketitle

\begin{abstract}
Cloud platforms have become essential in rapidly deploying application systems online to serve large numbers of users. Resource estimation and workload forecasting are critical in cloud data centers. Complexity in the cloud provider environment due to varying numbers of virtual machines introduces high variability in workloads and resource usage, making resource predictions problematic using state-of-the-art models that fail to deal with nonlinear characteristics. \\

Estimating and predicting the resource metrics of cloud systems across packet networks influenced by unknown external dynamics is a task affected by high measurement noise and variance. An ideal solution to these problems is the Kalman filter, a variance-minimizing estimator used for system state estimation and efficient low latency system state prediction. Kalman filters are optimal estimators for highly variable data with Gaussian state space characteristics such as internet workloads. \\

This work provides a solution by making these contributions: i) it introduces and evaluates the Kalman filter-based model parameter prediction using principal component analysis and an attention mechanism for noisy cloud data, ii) evaluates the scheme on a Google Cloud benchmark comparing it to the state-of-the-art Bi-directional Grid Long Short-Term Memory network model on prediction tasks, iii) it applies these techniques to demonstrate the accuracy and stability improvements on a realtime messaging system auto-scaler in Apache Kafka. The new scheme improves prediction accuracy by $37\%$ over state-of-the-art Kalman filters in noisy signal prediction tasks. It reduces the prediction error of the neural network model by over $40\%$. It is shown to improve Apache Kafka workload-based scaling stability by $58\%$. \\

\end{abstract}

\section{Introduction}
Dynamic scaling of distributed applications in response to changing workloads is critical to minimize costs, optimize resource usage, and meet Service Level Agreements (SLA) \cite{1}. Resource estimation and workload forecasting are critical in cloud data centers \cite{2}. Determining accurate resource state is a key aspect of system scaling, hence the need for fast and accurate resource state estimation, with changing numbers of virtual machines in the cloud systems and time-varying workloads \cite{3}. These conditions make predictions problematic using state-of-the-art models that fail to deal with these nonlinear workloads and system dynamics \cite{4}. High workload variability affects the input-output stability of autoscaling techniques, where stability is concerned with producing consistent scaling action as output over multiple iterations under similar workload inputs \cite{5}. \\

The task of estimating the resource utilization of large cloud systems over the internet and across networks influenced by unknown external and internal dynamics is also affected by high measurement noise and extreme points \cite{6}. To tackle the problem of estimation under the three conditions of high variability, noise and extreme points, an ideal solution is the Kalman filter, a variance-minimizing estimator used for system state estimation with noisy data \cite{7}, that can perform efficient low latency system state prediction \cite{8}. Kalman filters are optimal resource estimators when using highly variable data with a Gaussian noise such as internet-based workloads \cite{9}. \\

Compared to deep neural networks and machine learning methods, the Kalman filter is optimally resistant to noise and suitable for tasks involving low-latency state estimation \cite{10}. It has been demonstrated that state estimation under conditions of changing system dynamics is a problem well suited to the Kalman filter, a nonlinear state estimator for dynamic system models \cite{11}, which has shown effectiveness for estimation tasks in domains such as spacecraft and satellite trajectory planning and control, battery design, GPS tracking, image processing, and autonomous robotics \cite{12} where low latency estimation is vital.  \\

The basic Kalman filter is usable in estimating linear systems. To estimate nonlinear systems, alternative techniques include the Extended Kalman Filter, Unscented Kalman filter, and Sigma Point Kalman filter algorithms. The key goal of this work is to extend the nonlinear Kalman filter to incorporate Principal Component Analysis (PCA) \cite{13} and a novel attention model to increase robustness to noise. \\

This paper includes the following sections:
\begin{itemize}
    \item Background on the Kalman filter estimator
    \item Related work description of the state-of-the-art
    \item Detailed description of the approach
    \item Experimental evaluations
    \item Conclusions and recommendations
\end{itemize}

\section{Background}
Cloud providers are well served by using accurate resource estimates to ensure Quality of Service (QoS) to users while meeting SLA requirements enabling them to reduce costs and environmental footprint \cite{14}. \\

State-of-the-art time series prediction methods include nonlinear models such as Recurrent Neural Networks (RNN), Support Vector Machines (SVM) which are susceptible to high variability and noise \cite{15} \cite{16}. Moving average methods including Autoregressive Integrated Moving Average (ARIMA) are effective on high variability data, but cannot capture nonlinear system models and are only moderately effective with noisy data \cite{17}. The Kalman filter is suited to highly variable time series data such as network-based computing system metrics due to the capability to express both nonlinear state space and explicit noise models.

\subsection{Basic Kalman Filter} 
The Kalman filter uses observation data to estimate or predict the future system state. The basic Kalman filter consists of an observation model, which takes a state vector as input and computes a measurement vector using a linear operation, a state transition model which computes the next state given the current state vector, and noise models which add noise to the measurement and state vectors at each time step. When the next measurement is known, a correction is made to the state vector using the error between the known measurement and the computed measurement vector from the observation model. This state update step uses a linear operator called the Kalman gain taking the error as input and computing the state update vector as output. The following sets of equations capture these steps explicitly. \\

\subsubsection{Basic Kalman Filter Details}
The observations and observation noise vectors are represented by $z_{k}$ and $u_{k}$ respectively, while the state and state noise vectors are $x_{k}$, $v_{k}$ \& $w_{k}$ respectively \cite{11}. The observation equation is defined as 

\begin{equation}
 \label{eq:Z}
    z_{k} = Hx_{k} + u_{k-1}
\end{equation}

While the state update equation is defined as: 

\begin{equation}
 \label{eq:X}
    x_{k} = Ax_{k-1} + Bv_{k-1} + w_{k}
\end{equation}

$H$, $A$ \& $B$ are the observation, status transition, and status noise matrices respectively, which represent a Gaussian model. The state estimate from the state transition matrix at time $k-1$ is called the state prior $\hat x ^{-}_{k}$, and the state computed from the measurements at time step $k$ is called the state posterior $\hat x _{k}$ and is defined by the measurements at $k$. The error between the state transition estimate at time k and the states

\begin{equation}
\begin{aligned}
    & e^{-}_{k} = x_{k} - \hat x ^{-}_{k} \\
    & e_{k} = x_{k} - \hat x _{k}
\end{aligned}
\end{equation}

The covariance matrix P is calculated using this error, with prior and posterior values, and is in turn used in the calculation of Kalman gain K:

\begin{equation}
\begin{aligned}
 \label{eq:P1}
    & P^{-}_{k} = E(e^{-}_{k}.e^{-T}_{k}) \\
    & P_{k} = E(e_{k}.e^{T}_{k})
\end{aligned}
\end{equation}

With the initial values defined, the prior estimate error covariance matrix is updated 

\begin{equation}
\begin{aligned}
    & Q = E(w_{k}w^{T}_{k}) \\
    & P^{-}_{k} = AP_{k-1}A^{T} + Q
\end{aligned}
\end{equation}

The Kalman gain is defined 

\begin{equation}
\begin{aligned}
    & R = E(u_{k}u^{T}_{k}) \\
    & K_{k} = P^{-}_{k}H^{T}(HP^{-}_{k}H^{T} + R)^{-1}
\end{aligned}
\end{equation}

The updates to the state estimate and error covariance matrix are

\begin{equation}
\begin{aligned}
   & \hat x_{k-1} = \hat x^{-}_{k} +  K_{k}(z_{k}-H\hat x^{-}_{k}) \\
   & P_{k} = (I-K_{k}H)P^{-}_{k}
\end{aligned}
\end{equation}

\subsubsection{Basic Kalman Filter Summary}
The Kalman filter has the benefit of being noise tolerant and having easy-to-implement computations. It can be used in multi-object state estimation and requires a reasonable signal-to-noise ratio to be effective. The R and Q noise covariance matrices can cause divergence in the Kalman gain if they are inaccurate. \\

To enhance the robustness to noise, Principal Component Analysis (PCA) can be combined with Kalman filtering. In this scheme, the principal components of the observations are used as measurement variables, minimizing noise through dimensionality reduction in $R$. This leads to less variability in Kalman gain $K$ and more accurate estimate. \\ 

\subsection{Extended Kalman Filter} 
This work involves the estimation of nonlinear models hence the need for a nonlinear estimator such as the Extended Kalman Filter (EKF) \cite{18}. The measurement function and state function are nonlinear. The state transition equations are similar to the basic Kalman filter.  \\

\subsubsection{EKF Details}
The state transition and measurement equations of the nonlinear system are shown below:

\begin{equation}
\begin{aligned}
   & x(k) = f(x(k-1), v(k-1)) \\
   & z(k) = h(x(k), u(k))
\end{aligned}
\end{equation}

Where the function $f$ and $h$ are nonlinear functions. The nonlinear prior and posterior state update equations are

\begin{equation}
\begin{aligned}
   & \hat x^{-}_{k} = f(\hat x_{k-1}) \\
   & \hat x_{k} = \hat x^{-}_{k} + K_{k}(z_{k}-h(\hat x^{-}_{k}))
\end{aligned}
\end{equation}

\subsubsection{EKF Summary}
Kalman gain and covariance matrix update equations are unchanged from the basic Kalman filter algorithm. The EKF is optimal for situations where the observation and state variables have Gaussian white noise, and if they are independent of each other. The same criteria apply to the observation measurement $h$ and state transition function $f$. EKF has been used in road vehicle tracking and for spacecraft trajectory tracking and control \cite{18} showing its wide range of applications where low latency prediction is key. \\

The Unscented and Cubature Kalman filters are variants of the EKF for nonlinear models exploiting linearization techniques for structured data, while the Sequential Extended Kalman filter is a multimodal approach to the EKF. Principal component analysis can also be combined with each of these techniques to provide more robustness to noise and improve accuracy \cite{19}. These variants of the KF use sampling techniques to increase accuracy when using data with spatially correlated relationships. The intuition is to introduce attention as a generalized weighting technique for data with time-correlated data relationships. \\
 
\section{Related Work}
Measurements refer to inputs and states refer to outputs for common filtering algorithms. The nonlinear measurement and state model of the EKF include noise parameters that provide some robustness to extreme points. State-of-the-art methods for eliminating outliers including ARIMA \& gradient-based Savitzky-Gorlay filters either do not support nonlinear models or do not support independent measurement and state models for greater performance \cite{20}. Kalman filters do this inherently and are more robust when combined with other methods of mitigating extreme points such as weighting, point sampling, and principal component analysis.

\subsection{Unscented Kalman Filter}
The Unscented Kalman Filter (UKF) uses a sampling operation on nonlinear data, similar to the EKF \cite{21}. It uses the Kalman filter algorithm based on the Unscented Transform (UT) with deterministic sampling as the nonlinear observation function. The number of sampling points, also called Sigma-points, depends on the sampling strategy, often employed as the $2n + 1$ symmetrically-Sigma strategy, where $n$ refers to the number of samples. \\

\subsubsection{UKF Details}
The EKF uses linearization of the nonlinear function in order to make the state estimate at each step, which is accurate to the first order. To address this issue, the UKF algorithm uses sets of samples with weighted mean and covariance $x$ and $P_{x}$ respectively and makes estimates using nonlinear transformation of this state distribution $x_{i}$ and covariance $P_{x}$ in order to provide a more accurate estimate the up to the third order due the relatively small number of samples $n$. 

\begin{equation}
\begin{aligned}
     & x_{i} = \hat x_{k} + (\sqrt{(n+\lambda)P_{\hat x_{k}}})^{T}_{i}, i \in \{1,...,n\} \\
     & z^{-}_{k} = \sum_{i}h(x_{i}) \\
\end{aligned}
\end{equation}

$(\sqrt{(n+\lambda)P_{x_{k}}})^{T}_{i}$ refers to the $i^{th}$ row of the matrix square root. This transformation uses a set of weights whose mean is applied against measurement observations, and a set of parameter-determined weights whose covariance is multiplied against the measurement and status error covariances, with $\alpha$ describing the dispersion degree of samples, and $\beta$ describing the distribution of status samples, and $ \lambda = \alpha^{2}(k + n) - n $. 










\subsubsection{UKF Summary}
UKF is useful in cases where the system state vector is not known fully, using instead the observation function and partial distribution of the state vector, and more generally in cases where first order linearization leads to significant loss of accuracy in the EKF. UKF has been augmented to improve accuracy and speed, with more adaptive variants applied to the task of computer vision state estimation \cite{23}.  \\

\subsection{Sigma-Point Kalman filters} 
Spherical and radial measurement points which scale linearly with system state are best suited to the Sigma-point Kalman filter called the Cubature Kalman filter (CKF) used for estimating nonlinear system parameters \cite{24}. Similar in concept to the unscented Kalman filter, instead of random samples of measurement points at each time step, a set of spherical-radial cubature points is sampled. There are $2n$ of these cubature points of equal weight. The state update algorithm is as follows. \\

Initialize filter status $ \hat x_{0} = E[x_{0}] $ and covariance matrix $ \hat P_{0} = E[(x_{0}-\hat x_{0})(x_{0}+\hat x_{0})] $. Update the cubature points
\begin{equation}
\begin{aligned}
   & x_{i} = \sqrt{P_{\hat x_{k}}\lambda^{i}} + \hat x_{k}, i \in \{1,...,2n\} \\
   & \hat z^{-}_{k} = \frac{1}{2n}\sum(h(x_{i}))
\end{aligned}
\end{equation}

The Cubature filter is used in cases where the probability density around each measurement is either uniformly distributed or dispersed widely in all directions. To demonstrate this point, a study of the comparative performance of the EKF, UKF and CKF in the problem of tracking attitude sensors onboard the CBERS-2 satellite, containing high-frequency roll, pitch, and yaw data. The three algorithms had nearly identical accuracy. The EKF was the fastest of the three displaying the lowest time cost. The CKF was the only filter to satisfy the single operation duration limit, meeting the frequency requirements. \\

The main advantages of UKF and CKF were shown when initial conditions were corrupted with high noise-based error. In the case of 10\% initial error the EKF did not converge as fast, while a 20\% error cased the EKF to diverge for the entire experiment, while the UKF and CKF remained robust in non-ideal initial conditions. These initial condition effects are demonstrated in the approaches evaluated in this work. \\

\subsection{PCA \& Kalman Filters}
Principal component analysis (PCA) is used in many fields for data analysis. Along with feature extraction from data sets, PCA is also used for reducing noise and discovering hidden structure \cite{25}. Principal components refer to the most significant eigenvectors of the correlation matrix of any given set of column vector variables. The principal component scores refer to the eigenvalues. Low significance scoring principal components can be excluded from analysis in order to reduce noise and extract features.  \\

The Ensemble Kalman Filter with Principle Component Analysis uses PCA and clustering analysis to select the best model for inference \cite{26}. PCA is used for dimensionality reduction on the set of available models with the processed models then grouped by K-means clustering as a two-step sampling algorithm. Since the PCA is used as a selection method, the selected Kalman filter performs inference directly on the measurements using the standard Kalman filter algorithm. \\

The Enhanced Principal Component Analysis filter approach (EPCA) also combines PCA and Kalman filters \cite{27}. It uses Linear Predictive Coding (LPC) coefficients from input data whose PCA principal components are processed by Kalman filters as measurements. LPC is used in speech processing to encode signals at a low bit rate while maintaining quality. PCA and Kalman filters are used to reduce noise on input measurements increasing accuracy for state estimation. \\

EPCA is similar to the Extended Kalman filter PCA (EKF-PCA) algorithm used in this work, however, all these PCA-based Kalman filters lack the concept of \textit{attention} to account for the localized effects in the information distribution of input signals. Also, EPCA uses LPC to pre-process measurement data into PCA to determine the signal spectrum input gradient but the proposed attention Kalman filter uses raw measurements directly, reducing latency.  \\

\section{The Approach}
\subsection{Principal Component Analysis}



A good choice for principal components of a dataset matrix $X$ are the eigenvectors of the covariance matrix of $X$, which captures relationships between the columns of $X$. These are computed as



\begin{equation}
\begin{aligned}
    & pca(X) = \{\hat v_{\lambda_{max(\hat{D})}},...,\hat v_{\lambda_{min(\hat{D})}}\} : \hat{D} = \{\lambda_{i} \in D : \lambda_{i}>t\} 
\end{aligned}
\end{equation}

For threshold $t$, where $E$ is the expectation operator, columns $\{\hat v_{1},...,\hat v_{n}\}$ of $V$, and diagonal entries $\{\lambda_{1},...,\lambda_{n}\}$ of $D$ are the eigenvectors and eigenvalues of covariance matrix $C_{XX} = \frac{1}{n}[X - E[X]][X - E[X]]^{T}$ respectively. Dimensionality reduction is used to reduce noise in the data set by selecting only those eigenvalues whose eigenvectors are greater than $t$.

\subsection{Kalman Filter with Principal Component Analysis}
Two methods were considered to implement PCA with the Kalman filter: least squares method and linearization method. 

\subsubsection{Least squares method}
The KF-PCA with least squares algorithm involves first initializing the target observation vector as $z^{'}_{0} = pca(z_{0})$ where $z^{'}_{k} = pca(z_{k})$.
The Kalman filter observation projection matrix $h_{k-1}$ is updated using the least squares solution to the following matrix equation

\begin{equation}
    z^{'}_{0} = h_{0}x_{0}
\end{equation}

where $  z^{'}_{k-1} = h_{k-1}x_{k-1} $. Th observation vector is then calculated as 

\begin{equation}
   z^{-}_{k} =  h_{k-1}x^{-}_{k}
\end{equation}

This adjusted KF observation measurement is then used to estimate the next KF status $x_{k}$ using the basic KF prediction and update steps. \\

\subsubsection{Linearization method}
The KF-PCA with linearization algorithm involves updating the state as

\begin{equation}
    \hat x_{k} = x^{-}_{k}+ K_{k}( z^{'}_{k} - Hx^{-}_{k})
\end{equation}

$h(x^{-}_{k})$ is linearized to $Hx^{-}_{k}$ using the prediction Jacobian at $z^{'}_{k-2}$ and $\hat x_{k-1}$. \\

The linearization method produces lower prediction error on higher variability measurements than the least squares method. \\

KF-PCA with $t > 0$ has the benefit of noise reduction via PCA dimensionality reduction, which can be tuned to reduce noise by adjusting the PCA threshold parameter $t$. \\

\subsection{Attention}
Time series measurement data can have dependences at arbitrary distances in input sequences which can lower prediction accuracy when not incorporated into prediction models \cite{28}. This can happen where a subset of the EKF-PCA inputs is correlated to model state. Hence an attention network is added to locate these dependencies using weights on the input measurements of the EKF-PCA, produced by the attention network output layer $\hat{s}_{i}$ \cite{29}. \\

The attention network has an input layer $x^{i}$ which is weighted by layer weights $W$ and used to compute query $q^{i}$ and value $v^{i}$ layers.

\begin{equation}
\begin{aligned}
  & a^{i}=W^{a,i}x^{i} \\
 & q^{i}=W^{q,i}a^{i} \\
 & v^{i}=W^{v,i}a^{i} 
\end{aligned}
\end{equation}

This differs from the standard transformer attention layer in that it discards the key layer $ k^{i}=W^{k,i}a^{i} $, reducing the network size and complexity. The attention output is an accumulation of the value layer and softmax product $ b^{i}=\sum_{i}\hat{a}^{i}v^{i} $. \\

The add and normalization layer adds the attention input to the output and normalizes the result $\hat{b}^{i}=norm(x^{i} + b^{i})$. This then is connected to the linear layer $l^{i}=W^{i}\hat{b}^{i}$, which is connected to a softmax layer $ s_{i}=\frac{l^{i}}{\sum_{i}l^{i}} $. 
The softmax values are used with the measurement batch to compute the EKF input. 

\begin{equation}
 z_{k}=\sum^{n}_{i}{s}_{i}z_{i}
\end{equation}

This setup produces a set of weights applied to the EKF-PCA input values, as a type of sigma-point sampling. \\

\begin{figure}[h]
\includegraphics[trim=0 0 0 0,clip,width=0.9\linewidth, height=5cm]{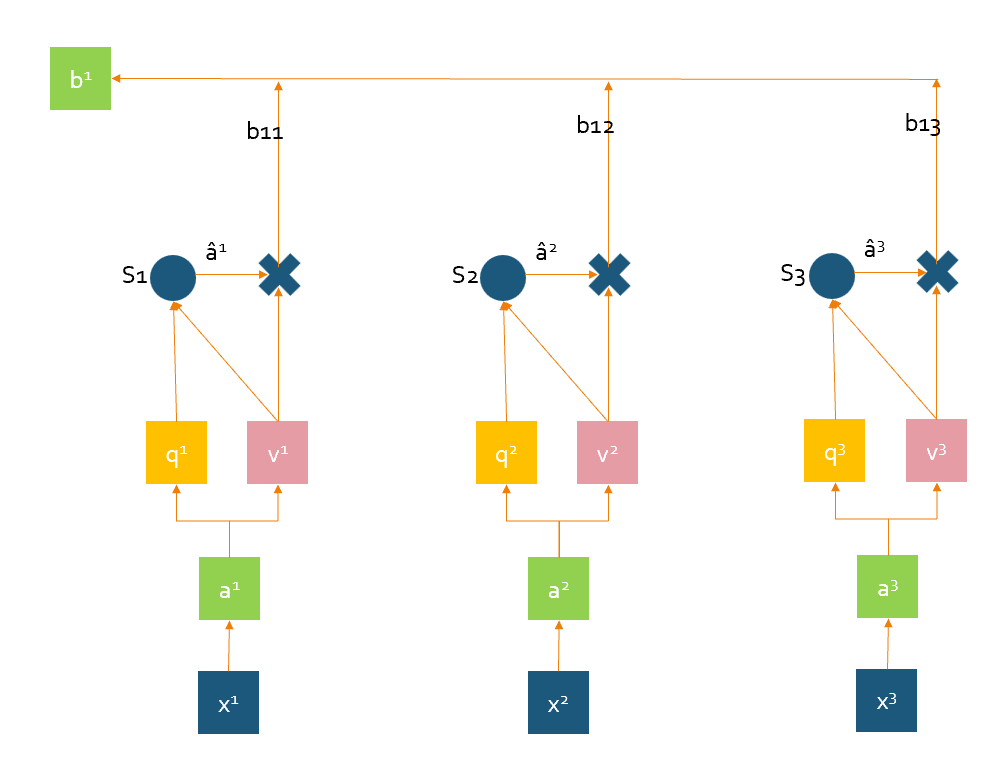} 
\caption{Attention Network. \\ N.B: $s_{i}$ Softmax layer}
\label{fig:fig0.1}
\end{figure}

\begin{figure}[h]
\includegraphics[trim=0 0 0 10,clip,width=0.9\linewidth, height=5cm]{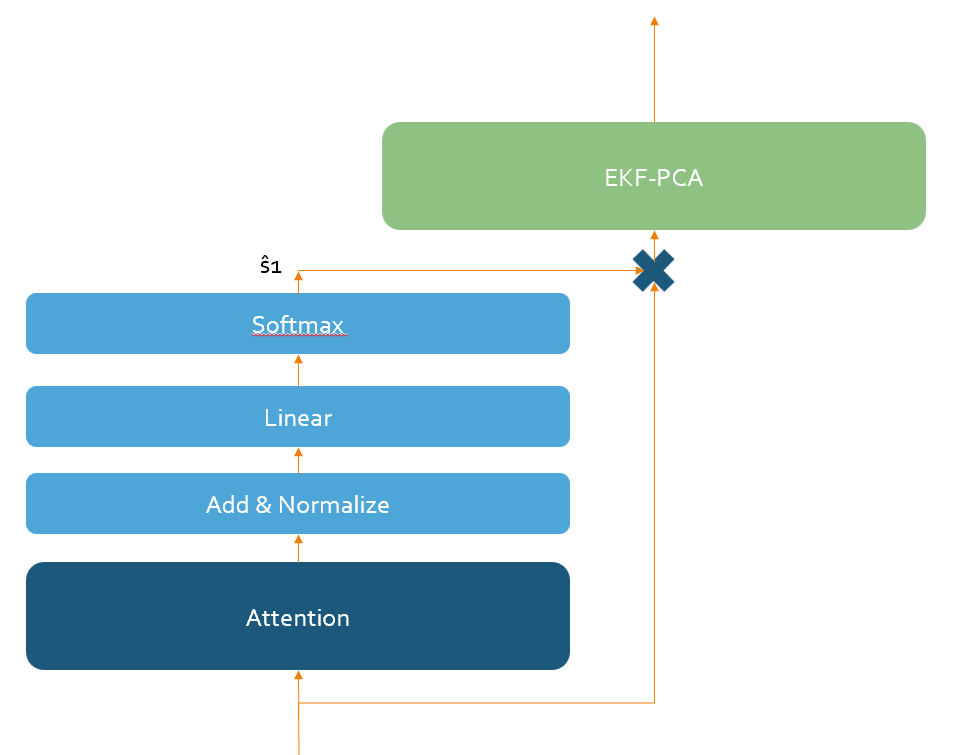}
\caption{AKF-PCA Estimator}
\label{fig:fig0.2}
\end{figure}

\section{Evaluations}

The evaluation consists of three parts: a set of simulated resource estimation experiments, a set of Google Cloud benchmark experiments, and a set of Apache Kafka real-time messaging experiments with simulated and real-world Twitter data using the Attention Kalman filter (AKF) and AKF with PCA (AKF-PCA) for estimation of future resource state.

\subsection{Resource Estimation with Simulated Data}
Resource estimation and prediction is a key part of provisioning distributed systems in response to request traffic, and is a key concern in cloud system management, such as for identifying anomalous workloads based on utilization patterns to mitigate ransomware and cryptographic attacks \cite{30}. In the simulation experiments, the estimators are applied to a set of resource estimation tasks and the results are analyzed to evaluate the performance of the AKF-PCA estimator. \\

\subsubsection{Estimated Resources and Estimator Types}

In this resource estimation effort, five types of Kalman filter are compared to each other in estimating the values of three types of resources. The Kalman filter types include the EKF, UKF, EKF with PCA (EKF-PCA), the UKF with PCA (UKF-PCA) and AKF-PCA. Joint estimators are designed by combining the nonlinear and PCA algorithms, following the pattern of the joint EKF-PCA estimator. The joint EKF-PCA combines the EKF \& EKF-PCA using the following algorithm called \textit{$\epsilon$ correction}:

\begin{equation}
    \epsilon_{k,EKF} = z_{k}-H\hat x^{-}_{k,EKF}
\end{equation}

\begin{equation}
    \epsilon_{k,PCA} = z_{k}-H\hat x^{-}_{k,PCA}
\end{equation}

\begin{equation}
\begin{aligned}
  &  \hat x_{k,EKF} = \hat x^{-}_{k,EKF} +  K_{k}\epsilon_{k,EKF} \\
  &  \hat x_{k,PCA} = \hat x^{-}_{k,PCA} +  K_{k}\epsilon_{k,PCA} \\
\end{aligned}
\end{equation}

\[
  x_{k} =  \{
\begin{array}{ll}
    x_{k,EKF}, & |\epsilon_{k,EKF}| \leq |\epsilon_{k,PCA}| \\
    x_{k,PCA}, & |\epsilon_{k,EKF}| > |\epsilon_{k,PCA}|  \\
\end{array}
\}
\]

In the first part of the evaluation, the estimation tasks include estimating single-step future values of a high variance CPU utilization trace from a cloud-based virtual machine running a messaging application, a low variance periodic Mackey-Glass 30 time series simulation for periodic signals with noise at SNR of 6dB \cite{31}, and to estimate the loss of a Grid Long Short Term Memory (LSTM) neural network trained using the mean squared error metric to aid in task scheduling. Task allocation for GPUs is useful for optimizing GPU utilization to improve quality of service \cite{32}. \\

\subsubsection{Estimation Results}
The Kalman filter priors and estimate plots are shown. Convergence latency can be measured using the difference in time between the occurrence of measurement values and the corresponding estimates for the estimators. Convergence latency affects the estimation performance of the UKF \& EKF with an average convergence latency of 9 intervals. The UKF-PCA \& EKF-PCA have convergence latency of 1 interval. 

\begin{figure}[h]
\begin{tikzpicture}
\node[inner sep=0pt] (thresholds) at (0,0)
    {\includegraphics[width=0.50\textwidth]{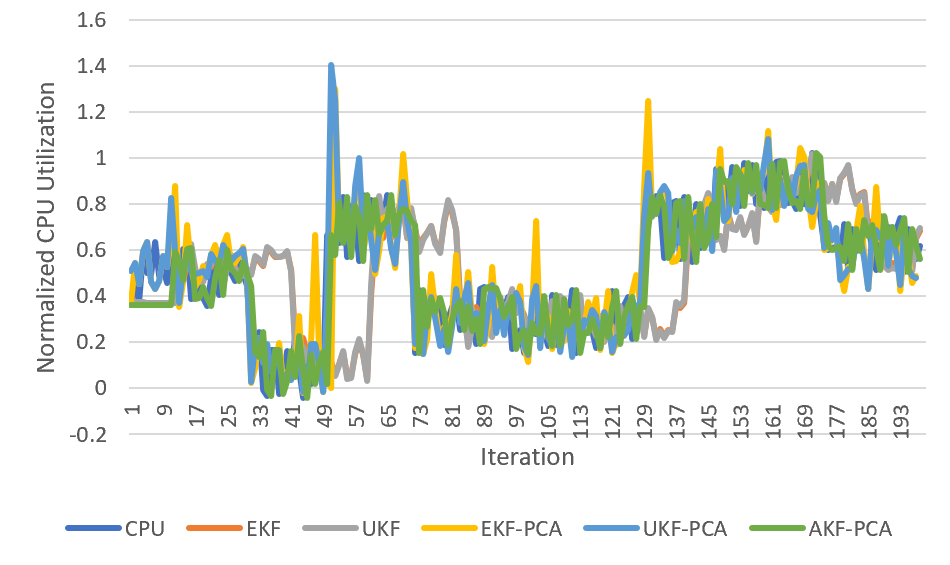}};
\draw[draw=red] (-1.45,-1.05) rectangle ++(0.50,2.95);
\draw[<->,thick,draw=red] (-1.55,2.15) -- (-0.85,2.15) node[midway,fill=white,text=black] {$e_{1}$};
\draw[<->,thick,draw=red] (-0.65,1.95) -- (-0.65,0.95) node[midway,fill=white,text=black] {$e_{2}$};
\end{tikzpicture}
\caption{CPU Utilization Estimation KF Priors}
\end{figure}
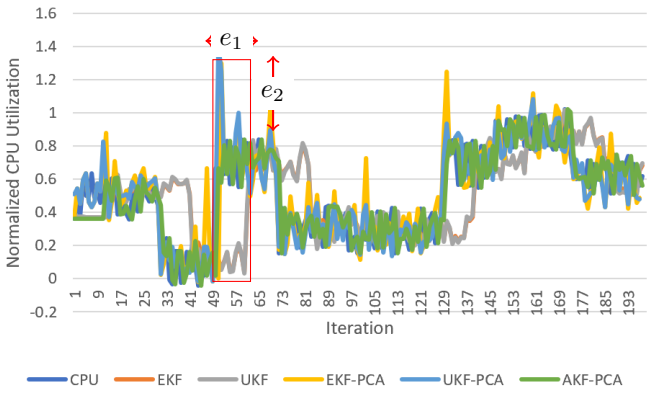

\begin{figure}[h]
\includegraphics[trim=0 0 50 50,clip,width=0.9\linewidth, height=5cm]{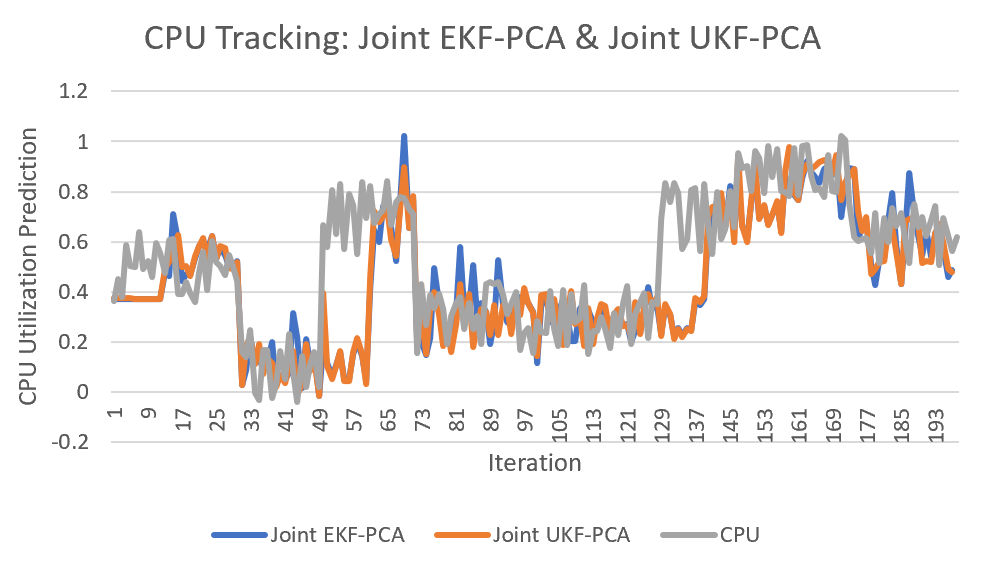} 
\caption{CPU Utilization Joint Estimator Priors}
\end{figure}

Compared to the EKF, UKF and KF, the EKF-PCA and UKF-PCA have better performance on the CPU utilization estimation task, where the EKF-PCA has a 28.2\% improvement in accuracy and 30.6\% reduction in error variance over the EKF. The EKF-PCA shows an even bigger improvement of 37.8\% in accuracy and 39.5\% reduction in error variance over the EKF. \\

Type $e_{1}$ error represents convergence latency, while type $e_{2}$ error represents peak divergence illustrated in the CPU estimation prior plots. $e_{1}$ error is pronounced in the UKF due to sigma point sampling for the unscented transform which makes it slower to respond to sharp changes in input values. $e_{1}$ errors significantly impact the joint UKF-PCA filter on the CPU utilization estimation task. \\

On the Mackey-Glass estimation task, the EKF-PCA and UKF-PCA show higher error rate than the EKF and UKF respectively. The joint EKF-PCA \& joint UKF-PCA perform well on this task however, with the joint EKF-PCA shows a 0.21\% improvement in standard deviation and 8.49\% reduction in error variance compared to the EKF, with similar error rate within 4\% for the joint UKF-PCA and the UKF. \\

The joint EKF-PCA and UKF-PCA error data is shown in the estimation error comparison. The largest reduction in error can be seen in the LSTM estimation task where EKF-PCA has a 86.4\% reduction in error variance and 79.2\% improvement in accuracy. The Joint EKF-PCA shows similar improvement at 81.4\% and 78.1\% respectively. In this task there is lower variance in the estimated signal than in the CPU and Mackey-Glass estimation tasks, indicating why the joint EKF-PCA and UKF-PCA have lower mean error due to PCA being a noise reduction method. 

\begin{figure}[h]
\includegraphics[trim=0 0 30 30,clip,width=0.9\linewidth, height=5cm]{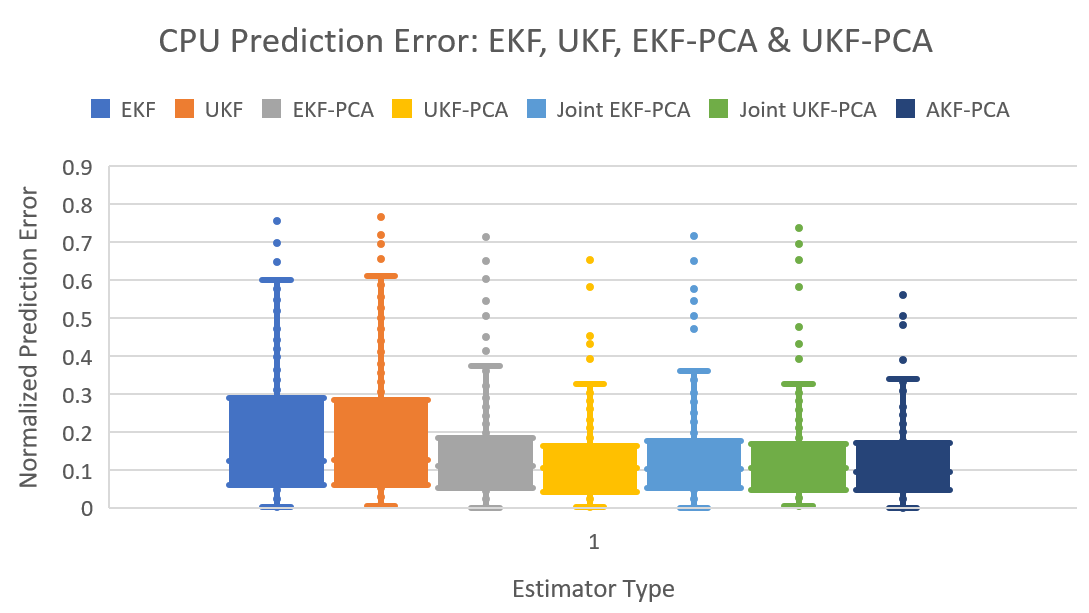} 
\caption{CPU Utilization Estimation Error}
\end{figure}

\begin{figure}[h]
\includegraphics[trim=0 0 30 30,clip,width=0.9\linewidth, height=5cm]{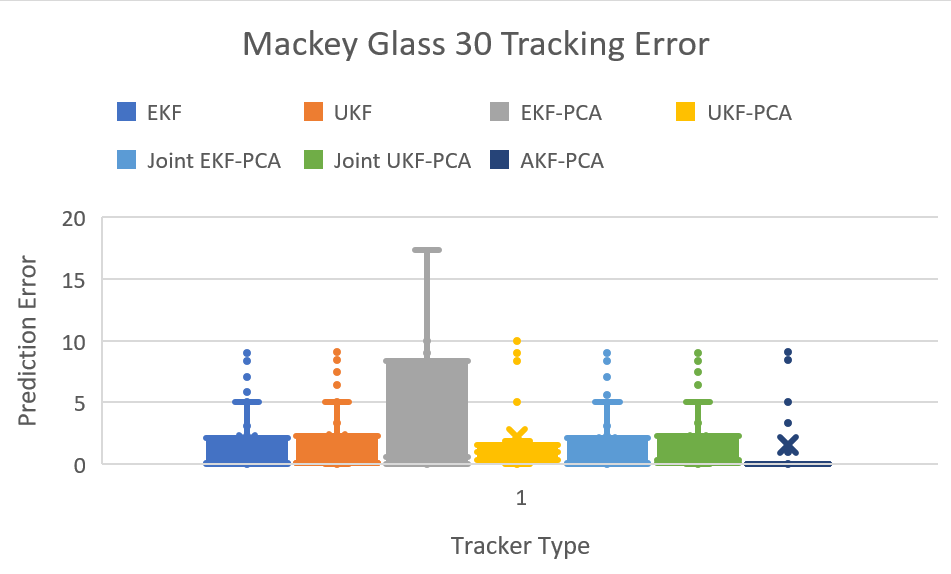}
\caption{Mackey-Glass 30 Time Series Estimation Error}

\label{fig:fig5}
\end{figure}

\subsubsection{Resource Estimation Experiments Discussion}

There is an overall improvement in accuracy for joint EKF-PCA of 29.9\%, 8.5\% and 78.1\% over the EKF, and 32.79\%, 4.7\% and 75.1\% for joint UKF-PCA over the UKF for CPU, Mackey-Glass and LSTM estimation tasks. However for high variance estimation tasks, such as for the CPU utilization estimation task the EKF-PCA has 4\% lower error than the EKF and UKF-PCA has 5\% lower error then the UKF. One conclusion is that the variance of the estimated signal is a useful criterion for selecting the estimation algorithm. \\

\begin{table}[t]
\centering
\begin{tabular}{ |p{1.0cm}|p{0.65cm}|p{0.65cm}|p{0.70cm}|p{0.70cm}|p{0.70cm}|p{0.70cm}| p{0.65cm}|  }
 \hline
 \multicolumn{8}{|c|}{Mean \& Variance for Estimator Errors (mean=$\nu$, std deviation=$\rho$)} \\
 \hline
 Signal & EKF  & UKF & EKF-PCA & UKF-PCA & Joint EKF-PCA & Joint UKF-PCA & AKF-PCA \\
 \hline
CPU $\nu$ & 0.1954 & 0.1973 & 0.1403 & 0.1246 & 0.1370 & 0.1326 & \textbf{0.1168}  \\
\hline
CPU $\rho$ & 0.1831 & 0.1839 & 0.1271 & 0.1112 & 0.1336 & 0.1327 & \textbf{0.0950} \\
\hline
Mackey-Glass 30 $\nu$ & 1.740 & 1.828 & 3.130 & 2.253 & \textbf{1.592} & 1.743 & 1.5971 \\
\hline
Mackey-Glass 30 $\rho$ & 2.748 & 2.867 & 3.961 & 3.407 & \textbf{2.742} & 2.870 & 3.414 \\
\hline
LSTM $\nu$ & 0.2380 & 0.2925 & 0.2072 & \textbf{0.2069} & 0.2532 & 0.2512 & 0.2094 \\
\hline
LSTM $\rho$ & 0.8511 & 0.9206 & \textbf{0.7634} & 0.7652 & 0.8684 & 0.8724 & 0.7838 \\
 \hline
\end{tabular}
\caption{Summary Statistics for Filter Estimator Errors}
 \label{tbl:tbl1}
 \end{table}

A rank matrix is computed for the estimation algorithms from the results table, where \[r_{j}=\frac{\sum_{i=0}^{n_{j}}r_{ij}}{n_{j}}\] represents the rank of each estimator. \\

The overall highest ranking estimation algorithm in terms of mean accuracy and standard deviation is AKF-PCA owing mainly to the suitability of the attention mechanism for modeling temporally distant dependencies in the simulated datasets \[rank(AKF\_PCA)=r_{6}=5.333\] where $rank(AKA\_PCA) > rank(UKF\_PCA) > rank(Joint\_EKF\_PCA) > rank(EKF\_PCA) > rank(EKF) > rank(Joint\_UKF\_PCA) > rank(UKF)$.
\\

\subsection{Google Cloud Benchmark Resource Estimation} 
An evaluation is done using the Google Cloud benchmark, to assess the performance of the different prediction approaches. The benchmark used is the Charlie memory trace which captures cloud CPU and memory utilization on Google Cloud containers \cite{33} from the processor instruction trace. The benchmark experiment setup includes state-of-the-art BiDirectional Grid LSTM with the Savitzky-Gorlay filter, and the best performing Kalman filters from the workload simulation experiments: the EKF, and the AKF \cite{4}. 

\subsubsection{Methodology}
To predict CPU utilization over a short time horizon, filtering algorithms are evaluated as direct predictors and as pre-filters for the Grid LSTM predictor. Comparisons are made regarding model accuracy, using high-variance workloads with outliers.

\begin{figure}[h]
\includegraphics[clip,width=1\linewidth]{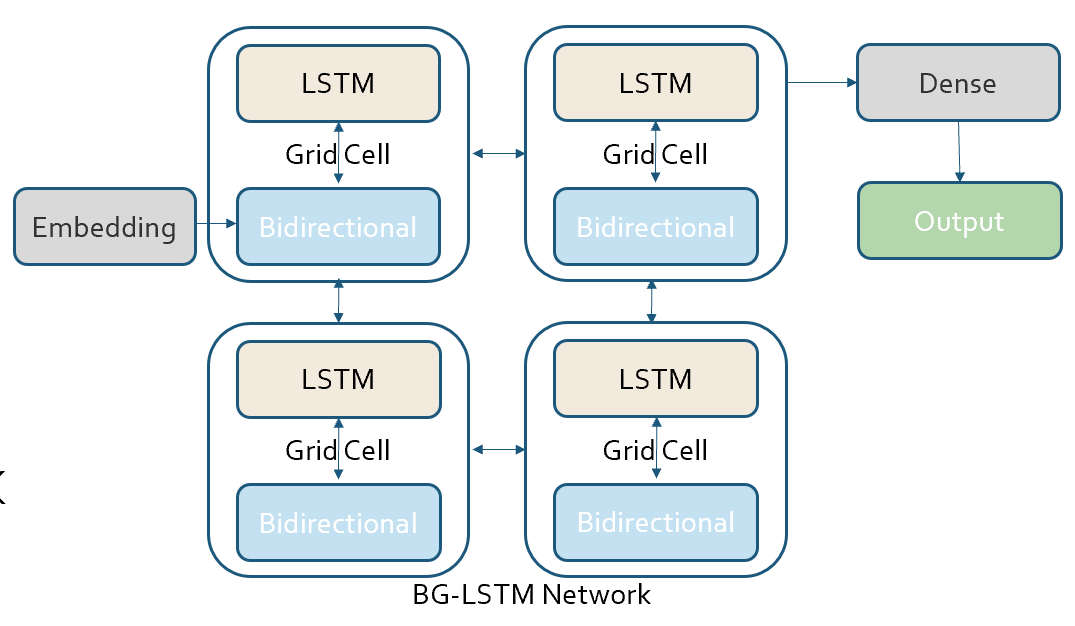}
\caption{Grid Bidirectional LSTM Network}
\label{fig:bglstm}
\end{figure}

\begin{figure}[h]
\includegraphics[clip,width=1\linewidth]{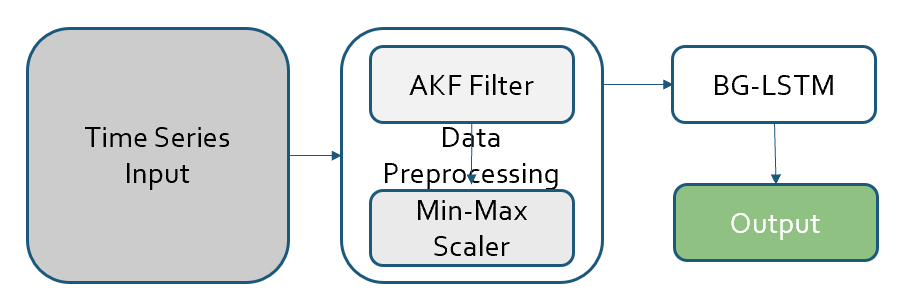}
\caption{Savitzky-Gorlay/Kalman filter \& Grid Bidirectional LSTM Setup}
\label{fig:filter_bglstm}
\end{figure}

\subsubsection{Results}
The performance metrics of the benchmark experiments include mean accuracy, error rate statistics, and prediction latency. For the Grid LSTM-based experiments, the number of epochs and training iterations is varied to characterize the impact on prediction accuracy. The results illustrate the superiority of the Extended Kalman filter and Attention Kalman filter in pre-filtering for the Grid LSTM algorithm (Grid) and as standalone estimators when compared to the Savitzky-Gorlay filter (Savgol). Results also show the superior accuracy of the Attention Kalman filter in single-step prediction compared to the Extended Kalman filter, the Savgol, and Grid algorithms.

\subsubsection{Grid LSTM Google Cloud Benchmark Experiments}

\begin{figure}
\begin{tikzpicture}
\node[inner sep=0pt] (grid_all) at (0,0)
    {\includegraphics[width=.25\textwidth]{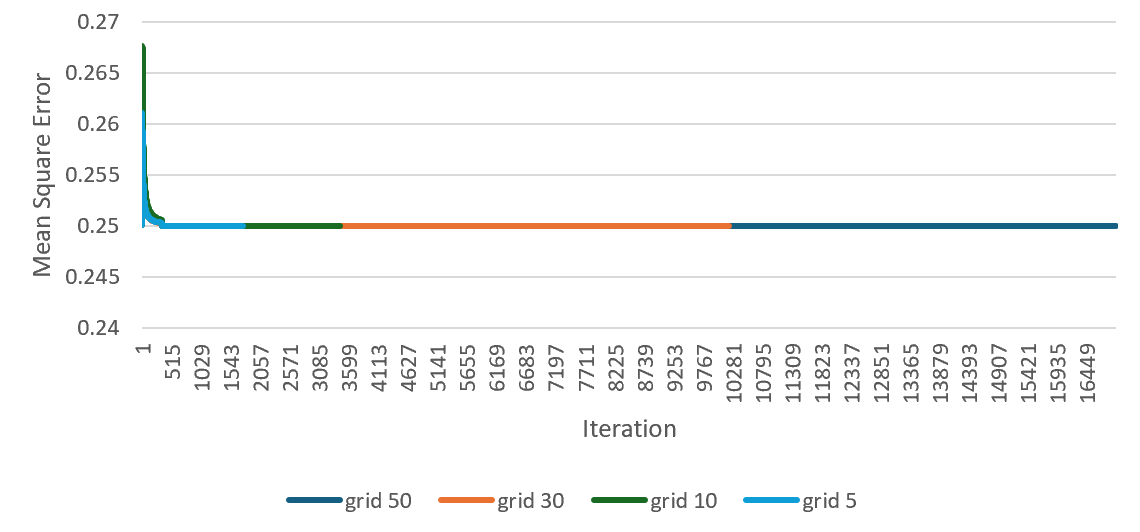}};
\node[inner sep=0pt] (grid_all_t1) at (4.3,0)
    {\includegraphics[width=.25\textwidth]{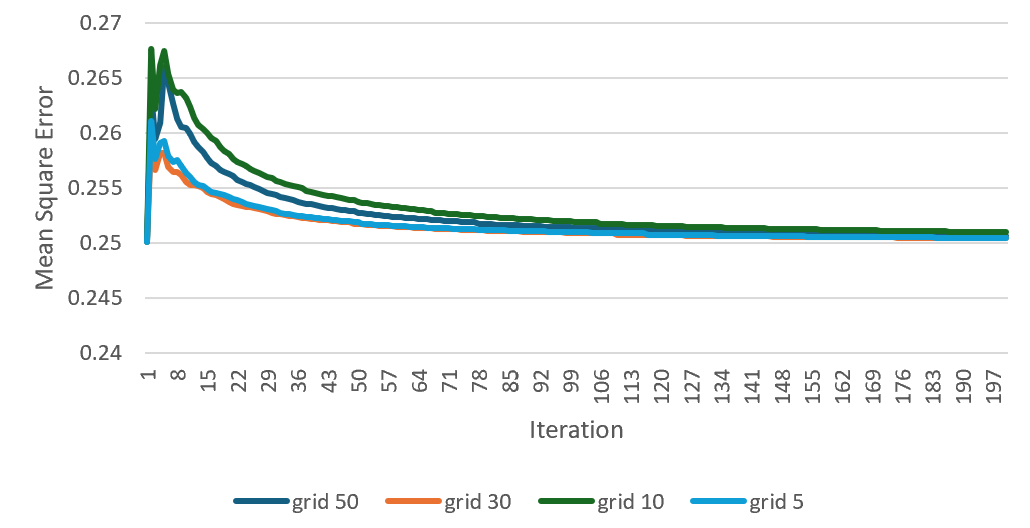}};
\node[inner sep=0pt] (grid_all_mean) at (4.3,-3.0)
    {\includegraphics[width=.25\textwidth]{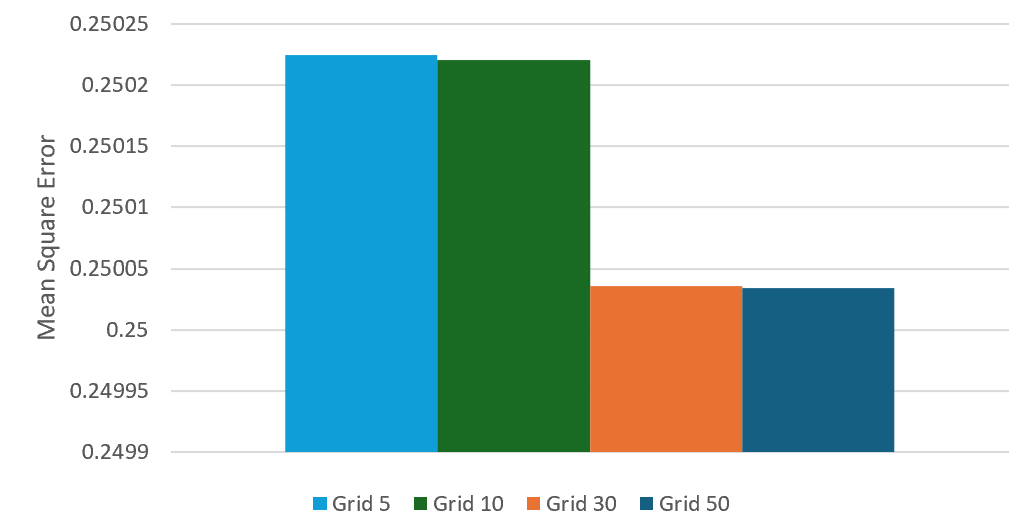}};
\draw[draw=red] (-1.75,0.1) rectangle ++(0.4,0.3);
\draw[-,thick,draw=red] (-1.36,0.3) -- (grid_all_t1.north west)
    node[midway,fill=white,text=red] {x16};\node[inner sep=0pt] (grid_all_t1) at (0,-3.0)
    {\includegraphics[width=.25\textwidth]{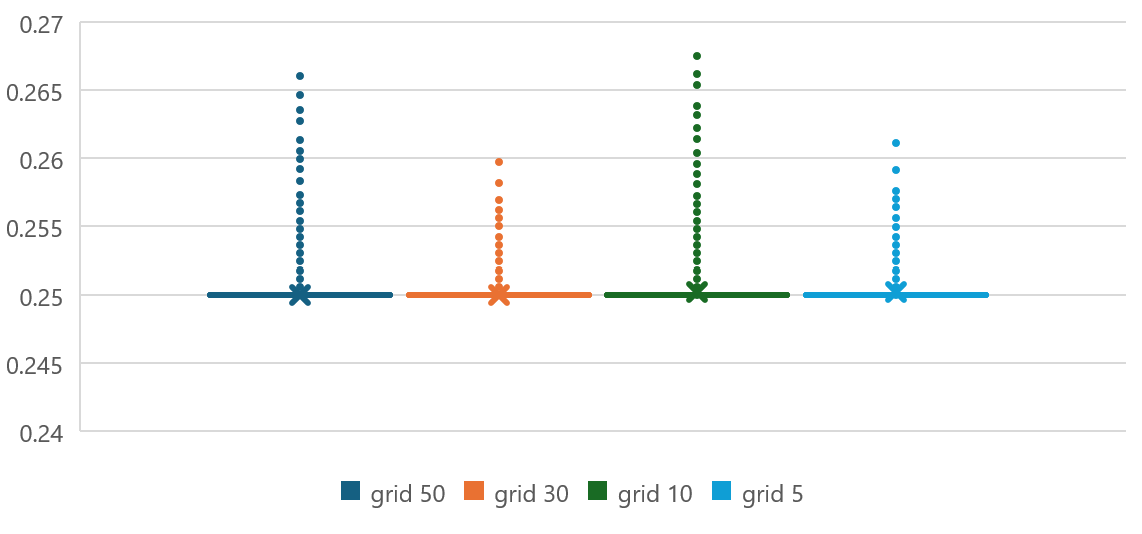}};
\end{tikzpicture}
\caption{Grid LSTM accuracy evaluation $n_{epoch} \in \{5,10,30,50\}$}
\end{figure}

The Grid LSTM Google Cloud benchmark experiment evaluates single-step CPU usage prediction accuracy. Evaluation and training are then repeated in a loop, where the training iterations have 5, 10, 30, and 50 epochs. There is a weak trend downward in cumulative mean squared error (MSE) as training epochs increase, with the epochs below 30 having mean squared error $\nu_{Grid30} = 0.250036$ and those above 30 having MSE $\nu_{Grid50} = 0.250034$.  The cumulative MSE over all epochs is $\nu_{Grid}=0.25013$.

\subsubsection{Savitzky-Gorlay Grid LSTM Google Cloud Benchmark Experiments}

\begin{figure}
\begin{tikzpicture}
\node[inner sep=0pt] (grid_all) at (0,0)
    {\includegraphics[width=.25\textwidth]{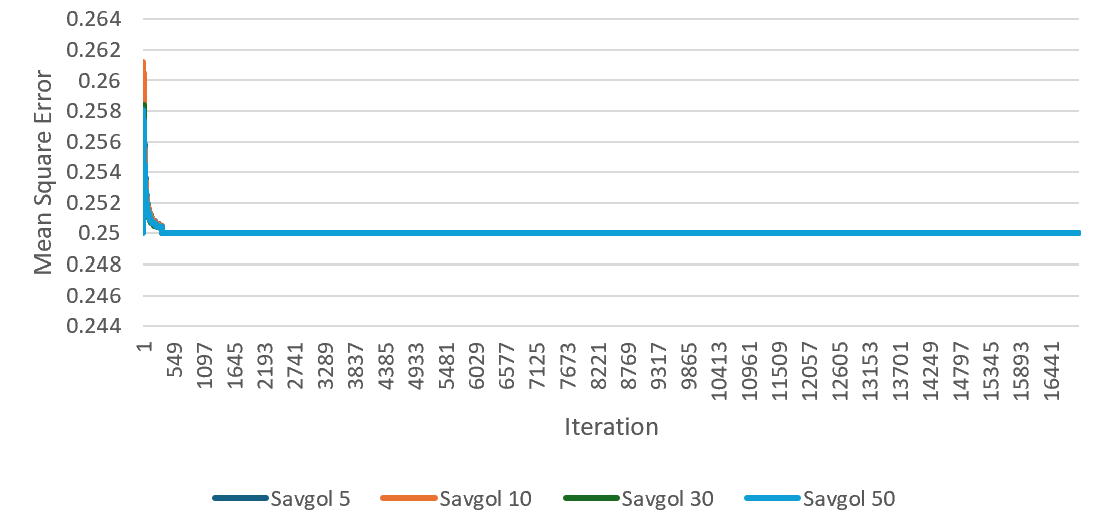}};
\node[inner sep=0pt] (grid_all_t1) at (4.3,0)
    {\includegraphics[width=.25\textwidth]{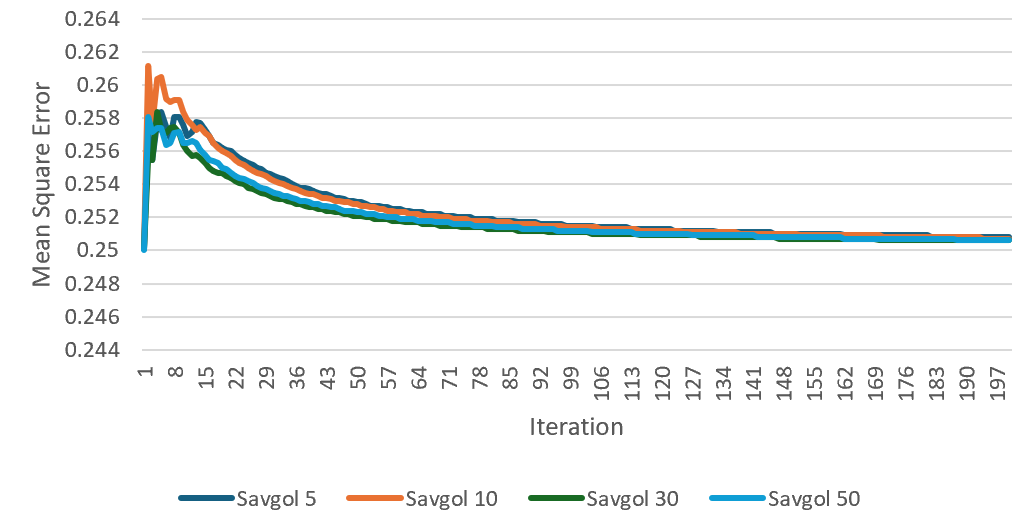}};
\node[inner sep=0pt] (grid_all_mean) at (4.3,-3.0)
    {\includegraphics[width=.25\textwidth]{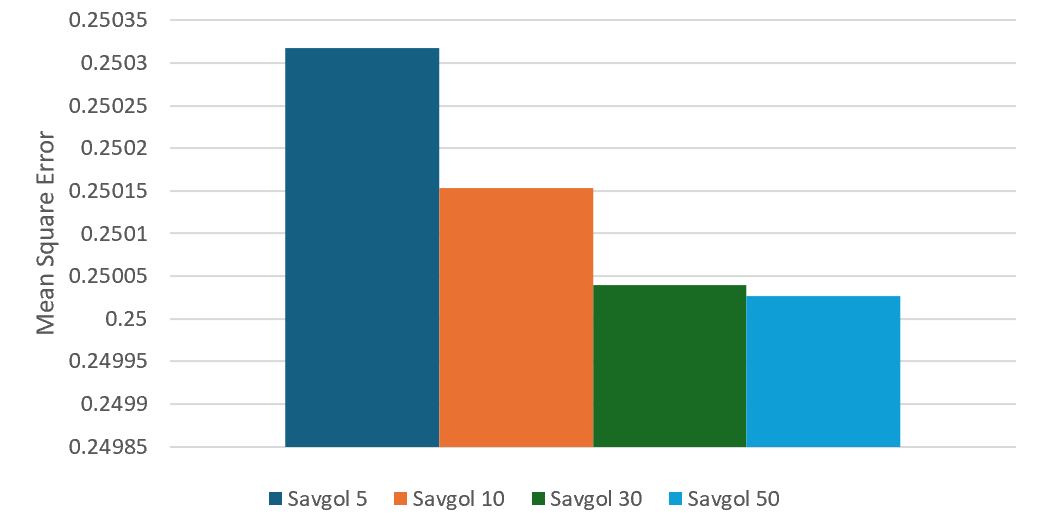}};
\draw[draw=red] (-1.75,0.1) rectangle ++(0.4,0.3);
\draw[-,thick,draw=red] (-1.36,0.3) -- (grid_all_t1.north west);
\node[inner sep=0pt] (grid_all_t1) at (0,-3.0)
    {\includegraphics[width=.25\textwidth]{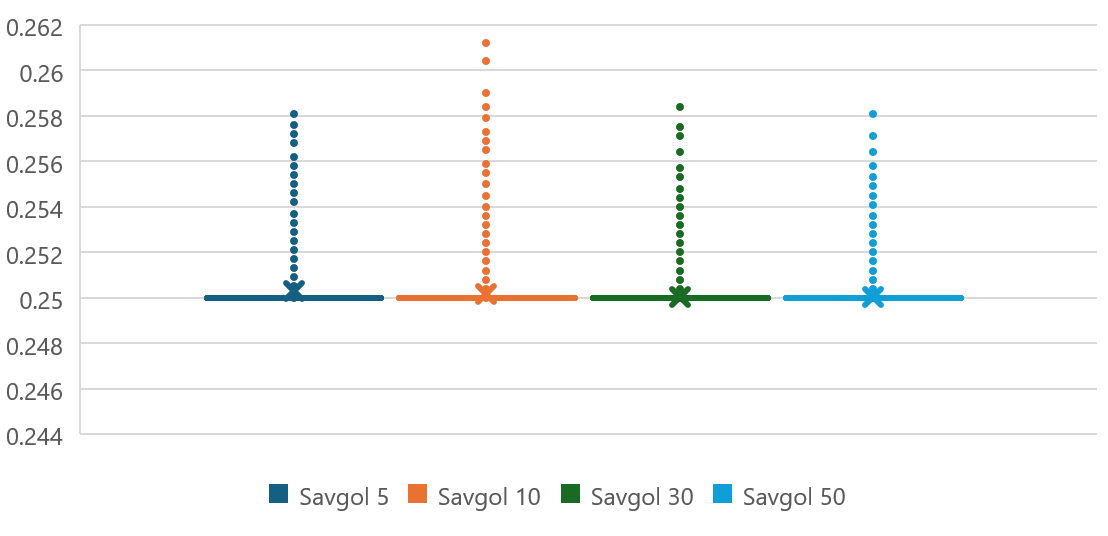}};
\end{tikzpicture}
\caption{Savitzky-Gorlay \& Grid LSTM accuracy evaluation $n_{epoch} \in \{5,10,30,50\}$}
\end{figure}

In the Savitzky-Gorlay (Savgol) filter \& Grid LSTM Google Cloud benchmark experiment, the CPU usage is smoothed out using the Savgol filter. The filtered single-step prediction accuracy is evaluated after training in iterations of 5, 10, 30 and 50 epochs. There appears to be a trend downward in MSE as training epochs increase, with the epochs below 30 having MSE $\nu_{Savgol30} = 0.25024$ and those above 30 having MSE $\nu_{SG50} = 0.250026$.  The MSE over all epochs is $\nu_{Savgol}=0.25003$. The average MSE for all epoch evaluations decreases monotonically $\rho_{Savgol5}=0.25032$, $\rho_{Savgol10}=0.25015$, $\rho_{Savgol30}=0.25004$, $\rho_{Savgol50}=0.25003$.

\subsubsection{Extended Kalman Filter Grid LSTM Google Cloud Benchmark Experiments}

\begin{figure}
\begin{tikzpicture}
\node[inner sep=0pt] (grid_all) at (0,0)
    {\includegraphics[width=.25\textwidth]{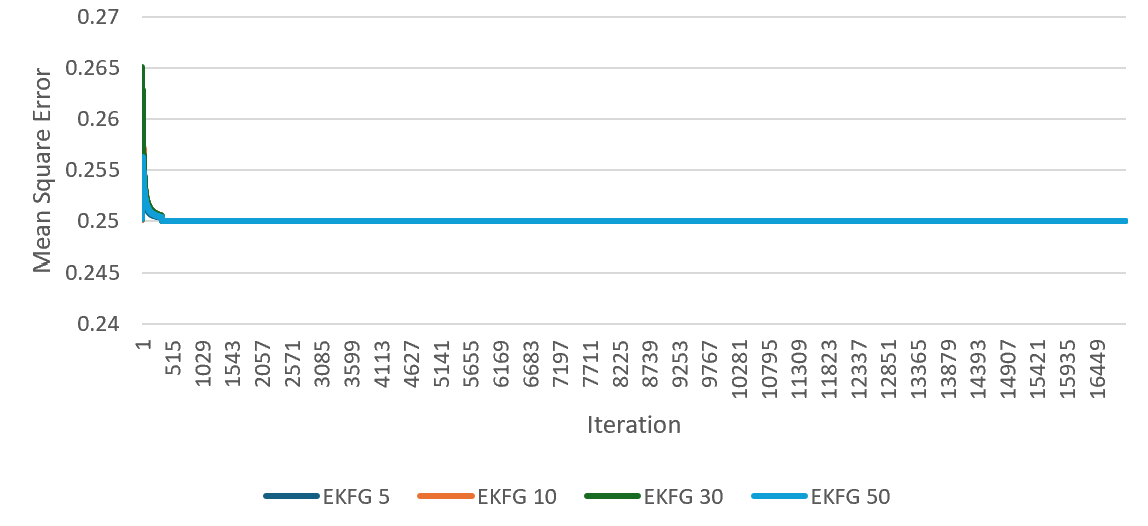}};
\node[inner sep=0pt] (grid_all_t1) at (4.3,0)
    {\includegraphics[width=.25\textwidth]{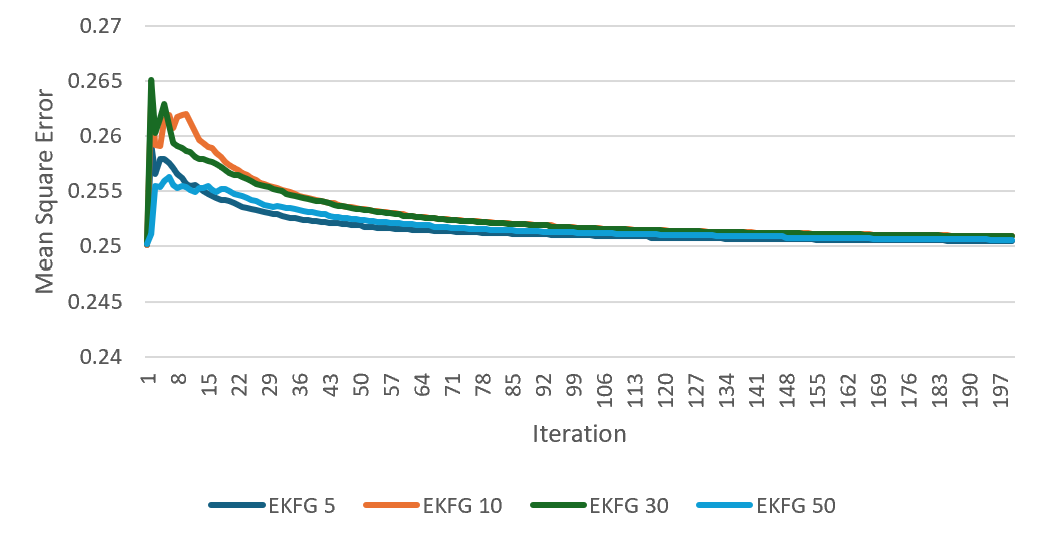}};
\node[inner sep=0pt] (grid_all_mean) at (4.3,-3.0)
    {\includegraphics[width=.25\textwidth]{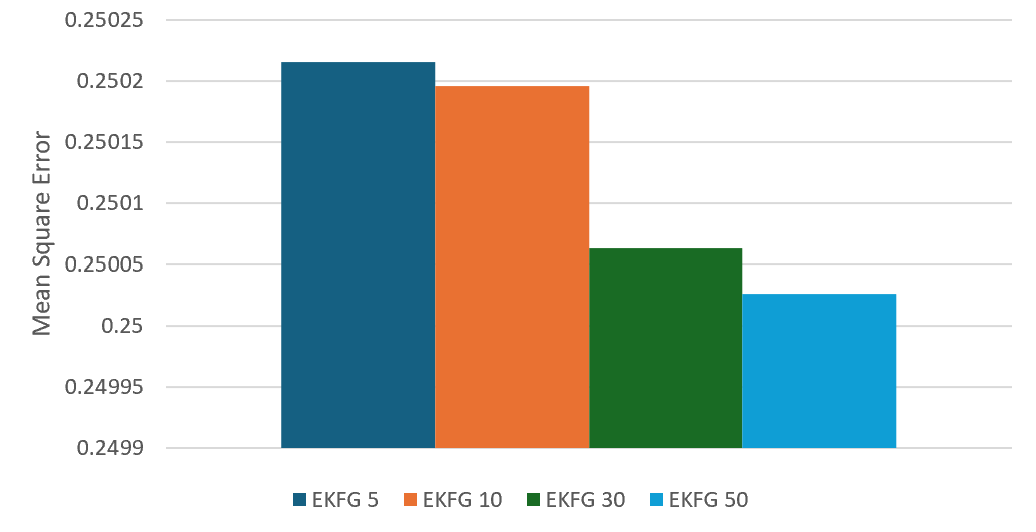}};
\draw[draw=red] (-1.75,0.1) rectangle ++(0.4,0.3);
\draw[-,thick,draw=red] (-1.36,0.3) -- (grid_all_t1.north west);
\node[inner sep=0pt] (grid_all_t1) at (0,-3.0)
    {\includegraphics[width=.25\textwidth]{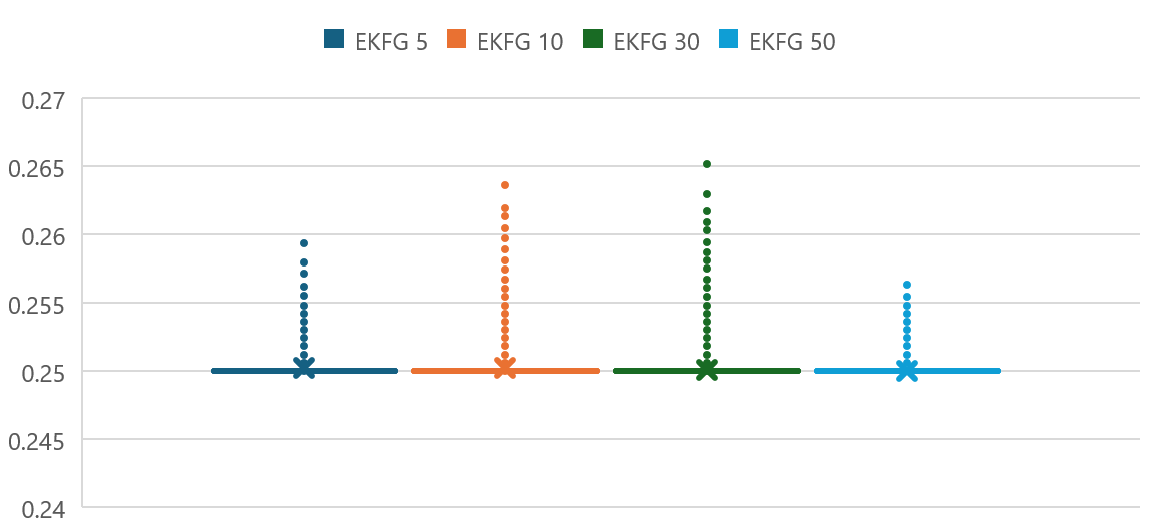}};
\end{tikzpicture}
\caption{Kalman (EKF) \& Grid LSTM accuracy evaluation $n_{epoch} \in \{5,10,30,50\}$}
\end{figure}

In the EKF \& Grid LSTM Google Cloud benchmark experiment, the CPU usage is pre-filtered using the EKF. The filtered single-step prediction accuracy is evaluated after training in iterations of 5, 10, 30 and 50 epochs.. There is a trend downward in cumulative MSE as training epochs increase, with the epochs below 30 having MSE $\nu_{EKF_G30} = 0.25021$ and those above 30 having MSE $\nu_{EKF_G50} = 0.25004$.  The cumulative MSE over all epochs is $\nu_{EKF_G}=0.25013$.

\subsubsection{Attention Kalman Filter Grid LSTM Google Cloud Benchmark Experiments}

\begin{figure}[h]
\begin{tikzpicture}
\node[inner sep=0pt] (grid_all) at (0,0)
    {\includegraphics[width=.25\textwidth]{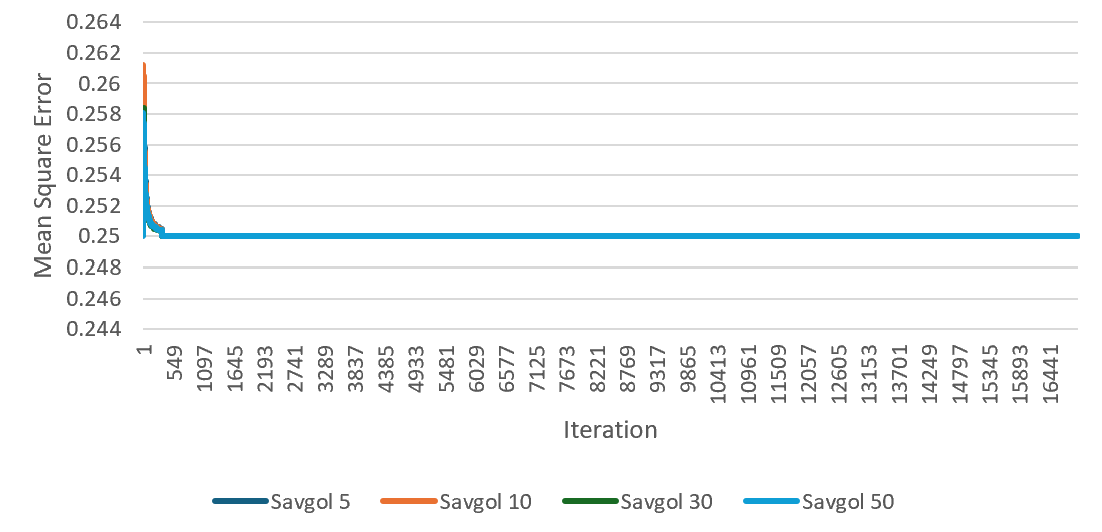}};
\node[inner sep=0pt] (grid_all_t1) at (4.3,0)
    {\includegraphics[width=.25\textwidth]{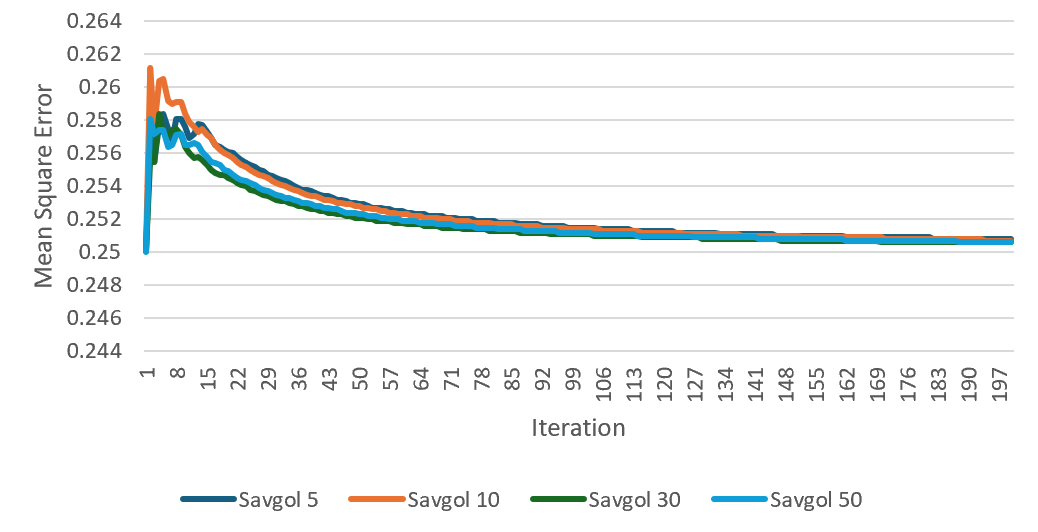}};
\node[inner sep=0pt] (grid_all_mean) at (4.3,-3.0)
    {\includegraphics[width=.25\textwidth]{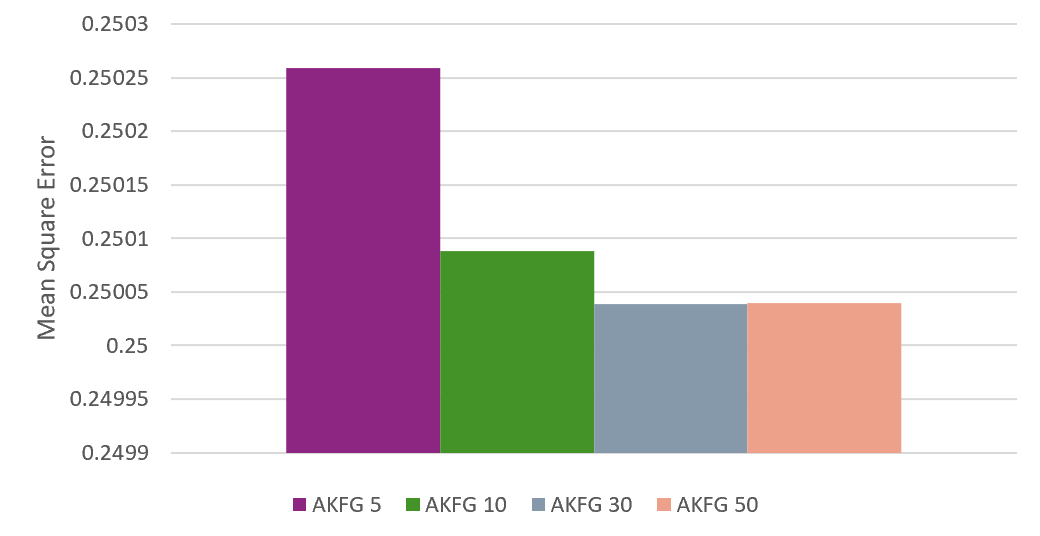}};
\draw[draw=red] (-1.75,0.1) rectangle ++(0.4,0.3);
\draw[-,thick,draw=red] (-1.36,0.3) -- (grid_all_t1.north west);
\node[inner sep=0pt] (grid_all_t1) at (0,-3.0)
    {\includegraphics[width=.25\textwidth]{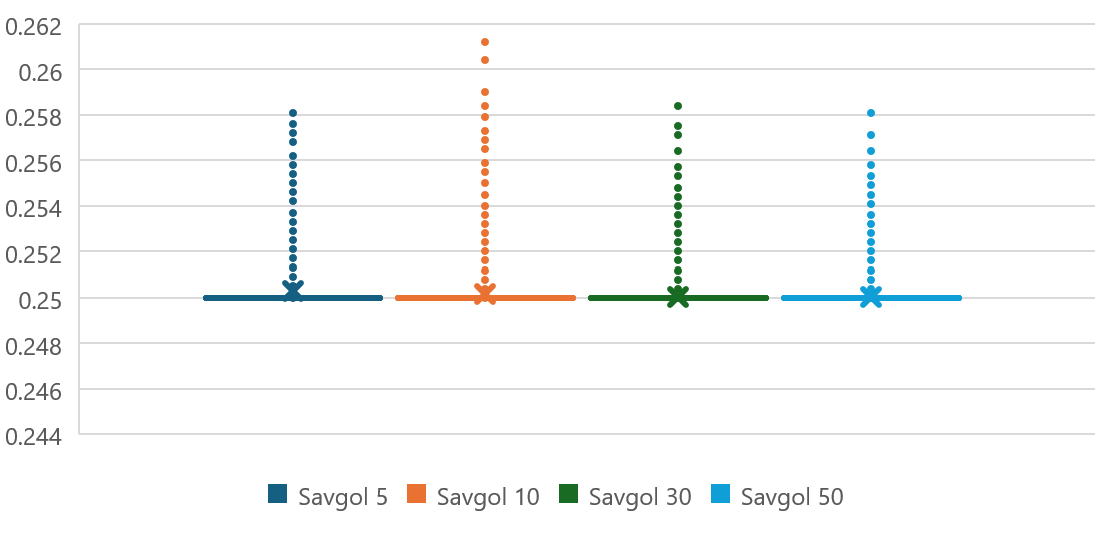}};
\end{tikzpicture}
\caption{AKF \& Grid LSTM accuracy evaluation $n_{epoch} \in \{5,10,30,50\}$}
\end{figure}

In the AKF \& Grid LSTM Google Cloud benchmark experiment, the CPU usage is pre-filtered by the AKF. The filtered single-step prediction accuracy is evaluated after training in iterations of 5, 10, 30 and 50 epochs. There is a trend downward in MSE as training epochs increase, with the epochs below 30 having MSE $\nu_{EKF_G30} = 0.25017$ and those above 30 having MSE $\nu_{EKF_G50} = 0.25004$.  The MSE over all epochs is $\nu_{EKF_G}=0.25011$, the best evaluation MSE of all the experiments.

\subsubsection{BGLSTM, Savitzky-Gorlay \& Extended Kalman Filter Google Cloud Benchmark Experiments}

\begin{figure}
\begin{tikzpicture}
\node[inner sep=0pt] (grid_all) at (0,0)
    {\includegraphics[width=.25\textwidth]{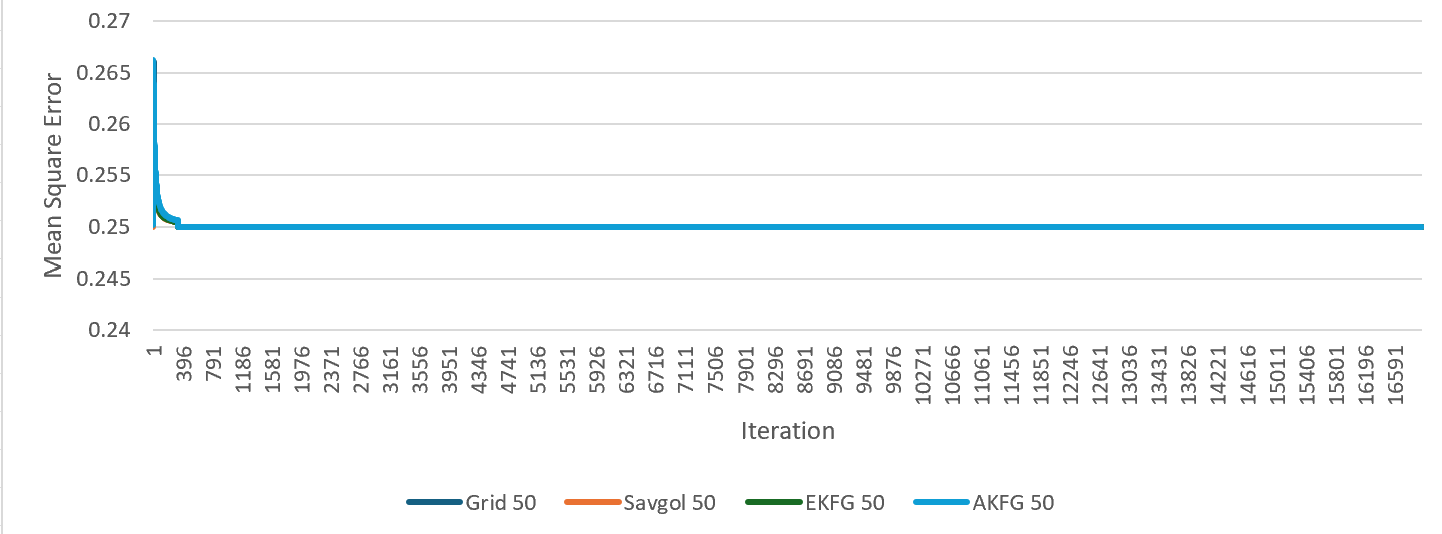}};
\node[inner sep=0pt] (grid_all_t1) at (4.3,0)
    {\includegraphics[width=.25\textwidth]{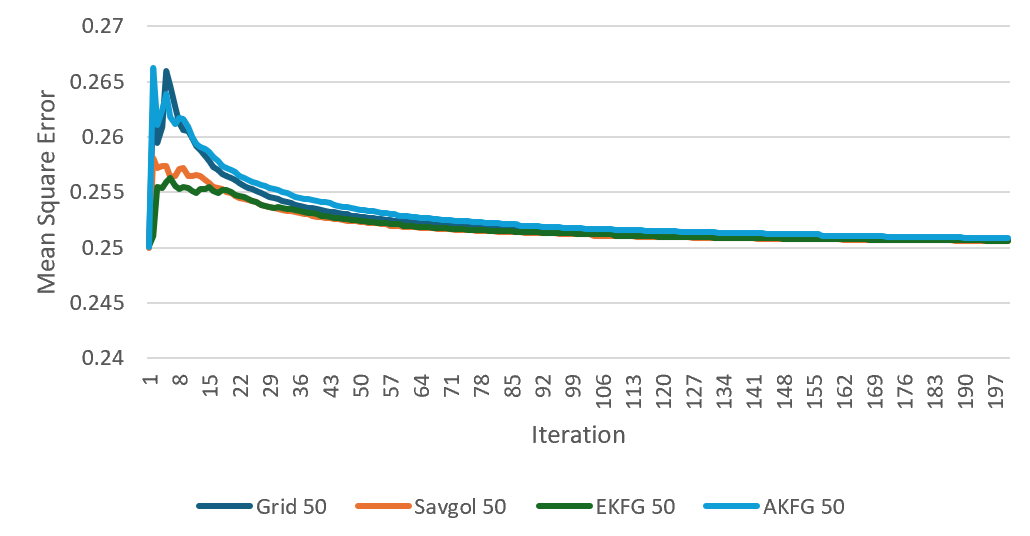}};
\node[inner sep=0pt] (grid_all_mean) at (4.3,-3.0)
    {\includegraphics[width=.25\textwidth]{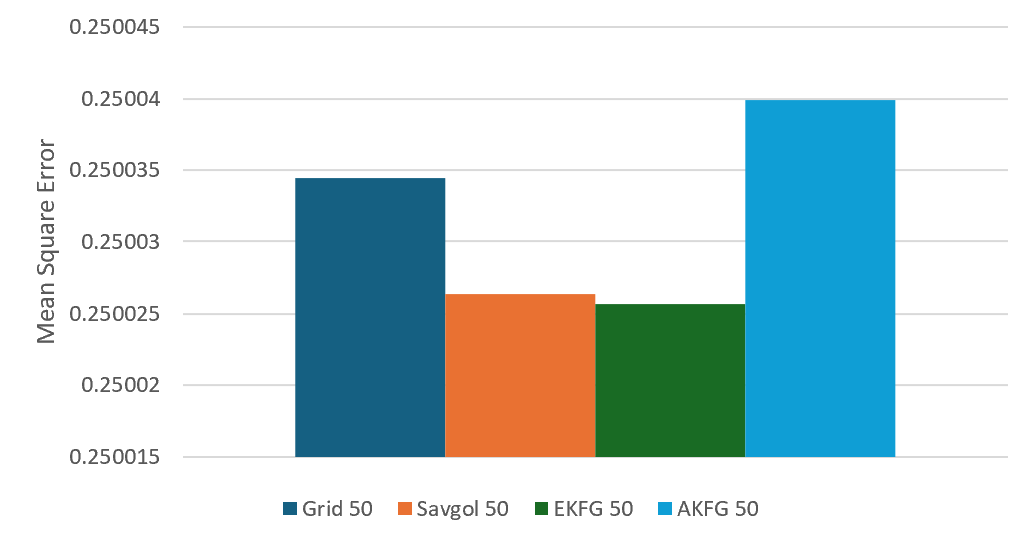}};
\draw[draw=red] (-1.75,0.1) rectangle ++(0.4,0.3);
\draw[-,thick,draw=red] (-1.36,0.3) -- (grid_all_t1.north west)
    node[midway,fill=white,text=red] {x16};\node[inner sep=0pt] (grid_all_t1) at (0,-3.0)
    {\includegraphics[width=.25\textwidth]{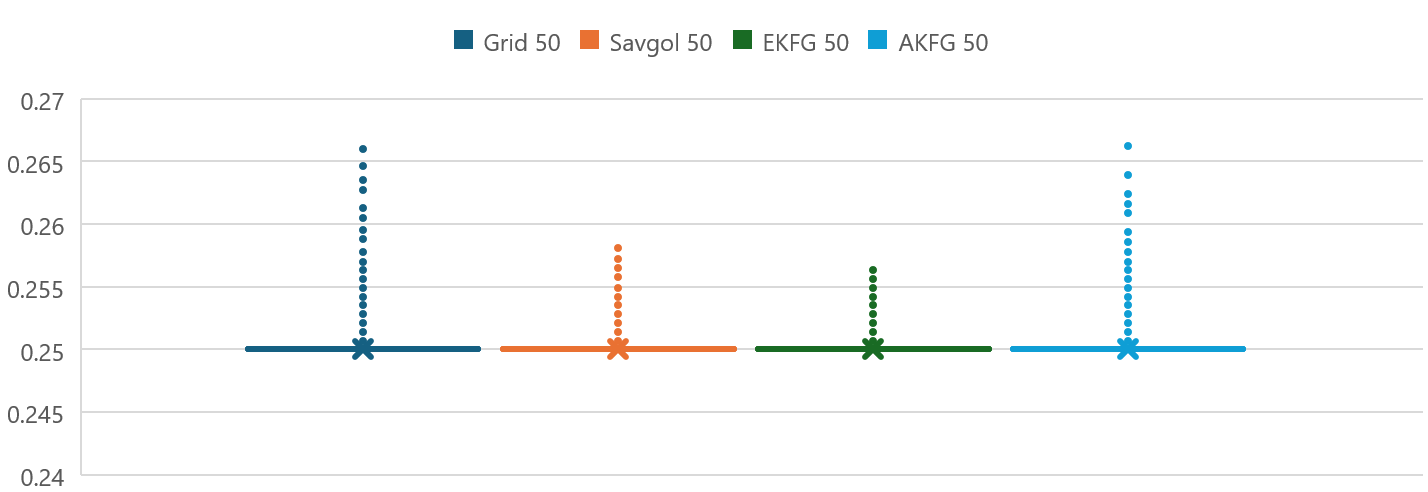}};
\end{tikzpicture}
\caption{Savitzky-Gorlay, Kalman \& Unfiltered Grid Bidirectional LSTM accuracy evaluation $n_{epoch} = 50$}
\end{figure}

In the EKF, AKF, Savitzky-Gorlay (Savgol) \& BGLSTM (Grid) Google Cloud benchmark experiment, the CPU usage is pre-filtered using the EKF in one setup, and the Savitzky-Gorlay filter in the second setup. The filtered and unfiltered single-step prediction accuracy is evaluated for all three experiments after training the Grid LSTM for 50 epochs. The AKFG is overfit at 50 epochs and has the highest MSE. The trend for average MSE is $\nu_{AKF_G} > \nu_{Grid} > \nu_{Savgol} > \nu_{EKF_G}$. 

\subsubsection{Extended Kalman Filter Google Cloud Benchmark Experiments}

\begin{figure}
\begin{tikzpicture}
\node[inner sep=0pt] (grid_all) at (0,0)
    {\includegraphics[width=.25\textwidth]{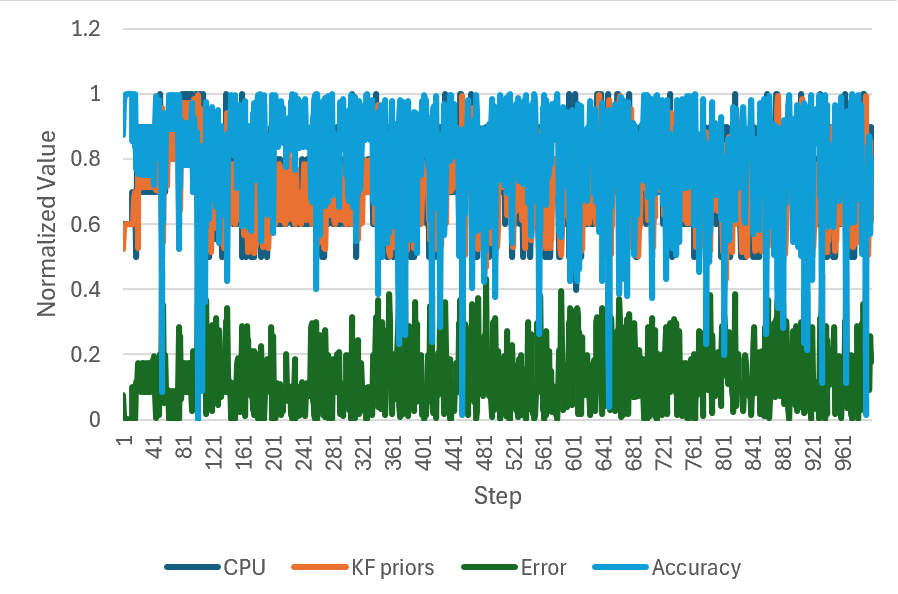}};
\node[inner sep=0pt] (grid_all_t1) at (4.3,0)
    {\includegraphics[width=.25\textwidth]{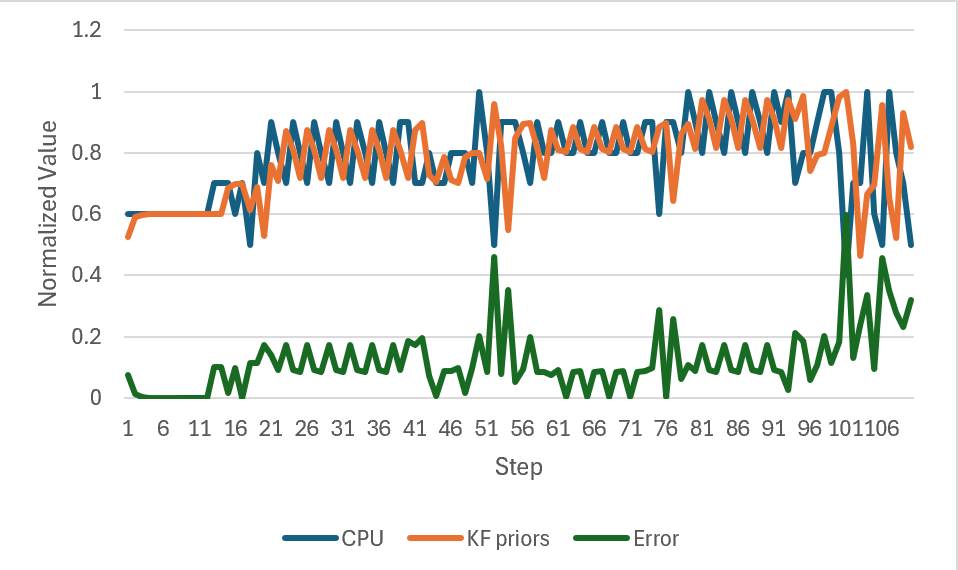}};
\draw[draw=red] (-1.75,-0.8) rectangle ++(0.4,1.9);
\draw[-,thick,draw=red] (-1.36,0.3) -- (grid_all_t1.north west);
\node[inner sep=0pt] (grid_all_t1) at (0,-3.0)
    {\includegraphics[width=.25\textwidth]{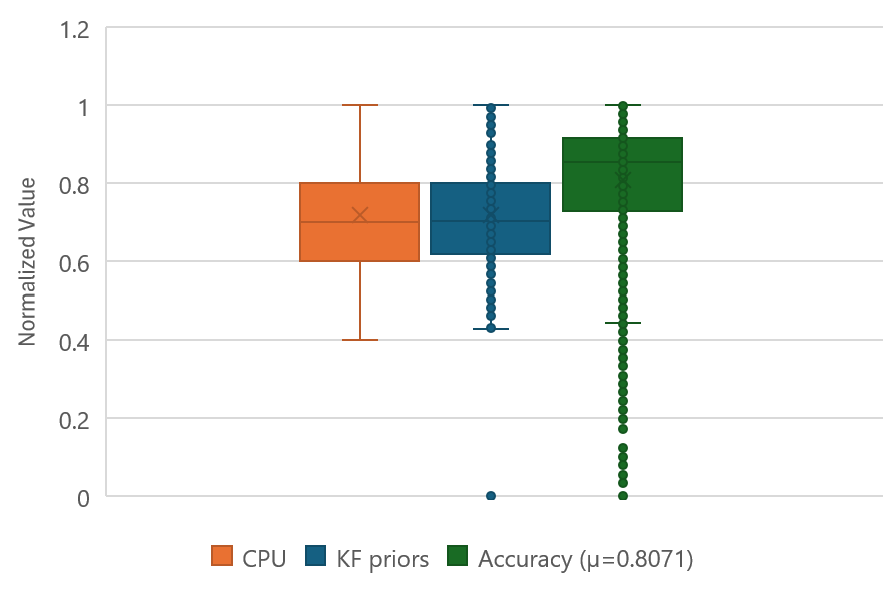}};
\end{tikzpicture}
\caption{EKF prediction accuracy - Google Cloud Benchmark}
\end{figure}

In the EKF Google Cloud benchmark experiments, single-step CPU usage predictions are made using the EKF from the trace data. There is a much higher accuracy than the Grid LSTM cumulative MSE $\epsilon_{EKF} = 0.8071$. There is less variance $\nu_{EKF} = 0.1175$ than the unfiltered CPU trace $\nu_{CPU} = 0.1333$. \\

\subsubsection{Attention Kalman Filter Google Cloud Benchmark Experiments}

\begin{figure}[h]
\begin{tikzpicture}
\node[inner sep=0pt] (grid_all) at (0,0)
    {\includegraphics[width=.25\textwidth]{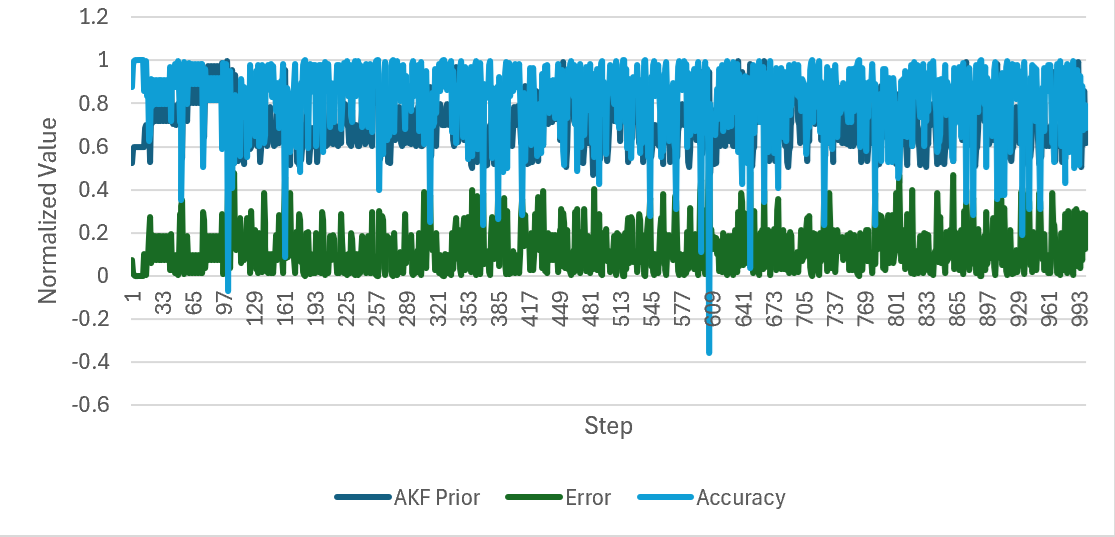}};
\node[inner sep=0pt] (grid_all_t1) at (4.3,0)
    {\includegraphics[width=.25\textwidth]{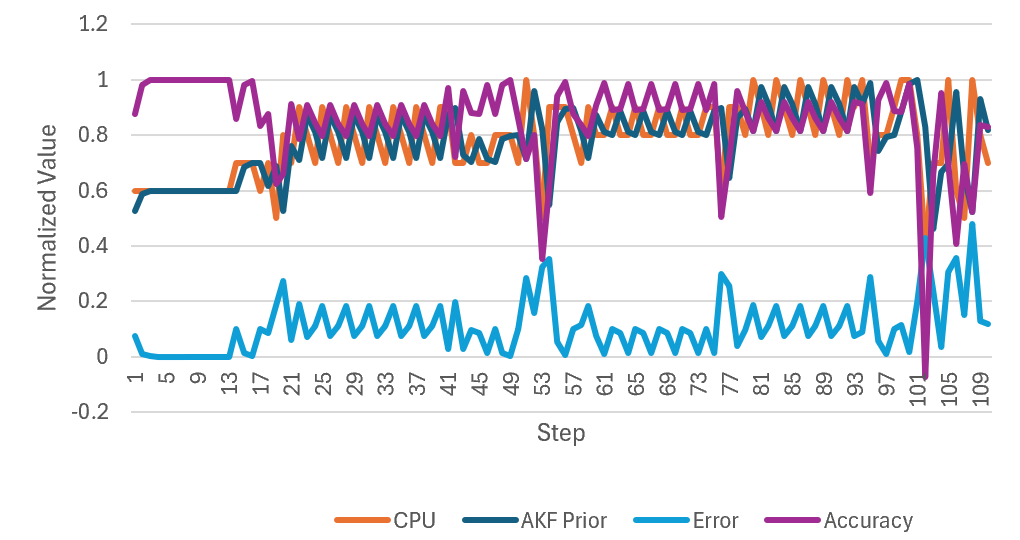}};
\draw[draw=red] (-1.75,-0.2) rectangle ++(0.4,1.1);
\draw[-,thick,draw=red] (-1.36,0.3) -- (grid_all_t1.north west);
\node[inner sep=0pt] (grid_all_t1) at (0,-3.0)
    {\includegraphics[width=.25\textwidth]{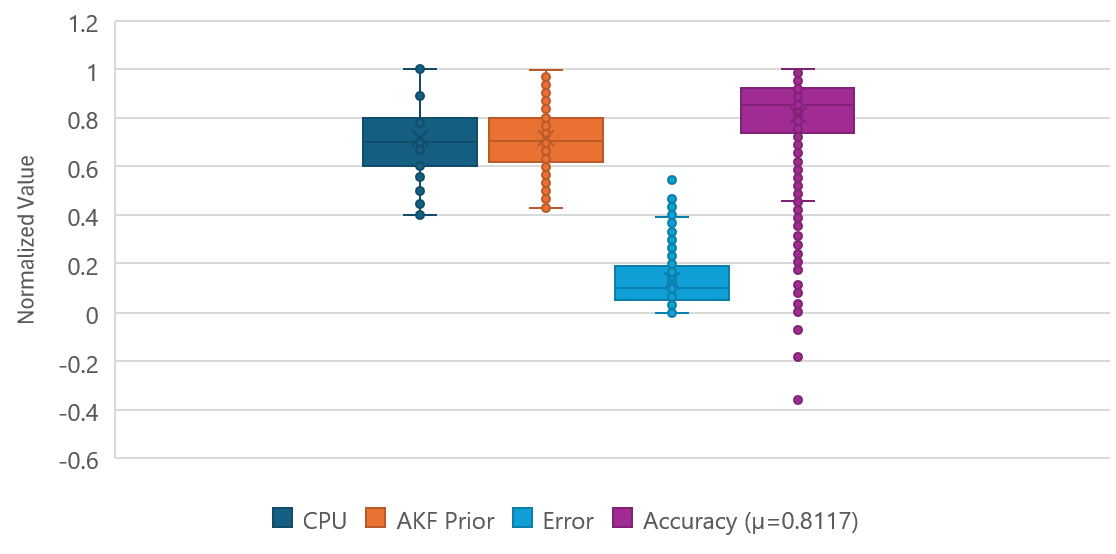}};
\end{tikzpicture}
\caption{AKF prediction accuracy - Google Cloud Benchmark}
\end{figure}

In the EKF Google Cloud benchmark experiments, single-step CPU usage predictions are made using the AKF from the trace data. There is less MSE variance $\nu_{AKF} = 0.1175$ than the unfiltered CPU trace.  There is also the lowest cumulative MSE in all the Google Cloud benchmark experiments $\epsilon_{AKF} = 0.8117$.

\subsubsection{PCA-based Attention $\&$  Extended Kalman Filter Benchmark Experiments}
The effectiveness of AKF-PCA and EKF-PCA as pre-filtering algorithms for the BGLSTM network is demonstrated in a series of experiments on the Google Cloud benchmark. The APCAG (AKF-PCA pre-filtered BGLSTM) and EPCAG (EKF-PCA pre-filtered Grid) produce improvements of $42.28\%$ and $42.21\%$ in cumulative RMSE over the Savitsky-Gorlay pre-filtered BGLSTM (Savgol) and the unfiltered BGLSTM (Grid). There is a significant reduction in cumulative MSE for APCAG and EPCAG when compared to Grid and Savgol algorithms, where $\nu_{Savgol} > \nu_{Grid} > \nu_{EPCAG} > \nu_{APCAG} > \nu_{EKFG}$. \\

\begin{figure}[h]
\includegraphics[trim=0 0 5 5,clip,width=0.9\linewidth, height=5cm]{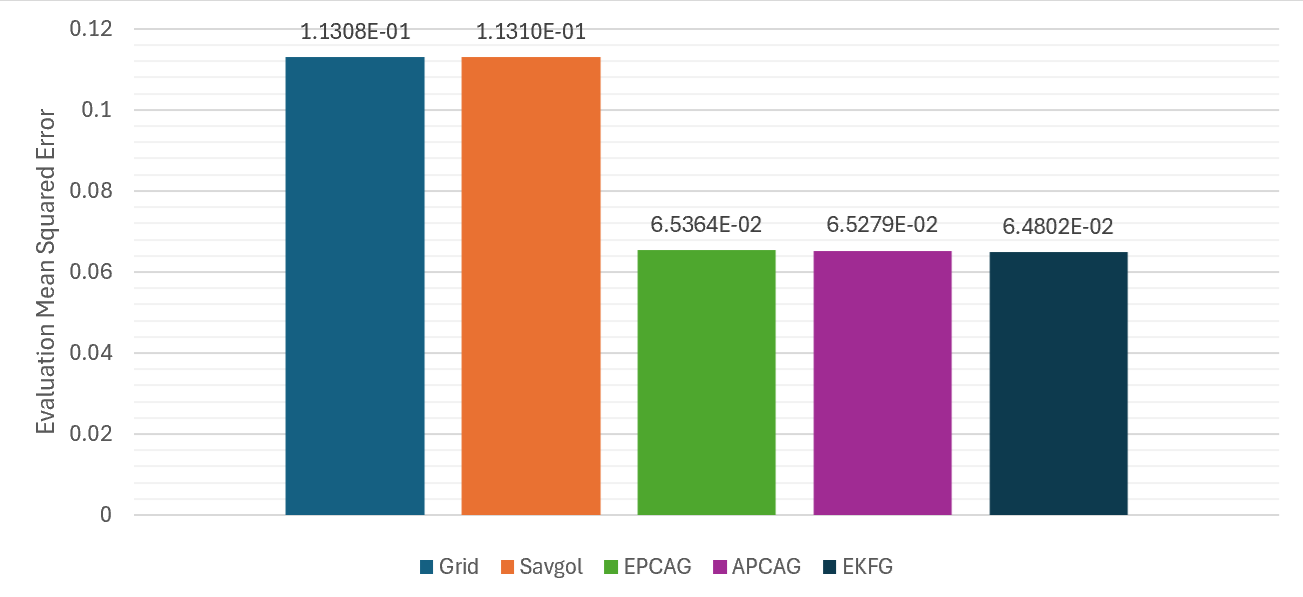} 
\caption{EKFG, EPCAG, APCAG, Savgol \& Grid All Epochs RMSE}
\end{figure}

 In this comparison, the PCA-based APCAG and EPCAG did not perform better than the EKFG but performed better than the Savgol and Grid algorithms due to the significant but limited impact of PCA noise reduction on the Google cloud CPU trace data. \\

\subsubsection{Google Cloud Benchmark Discussion}
The Grid algorithm is beneficial for capturing two-dimensional spatial dependencies in time series data. Input filters improve the accuracy of the Grid LSTM in all experiments because they mitigate extreme points and noise. The Savgol algorithm involves convolution with a $d=5$ term polynomial of degree $k=2$ to smooth out the inputs. The EKF-PCA and AKF-PCA algorithms had over $40\%$ lower RMSE than the Grid, Savgol algorithms, and similar RMSE to the EKFG and AKFG algorithms due to high variance in the benchmark data.  \\

\begin{figure}[h]
\includegraphics[trim=0 0 30 30,clip,width=0.9\linewidth, height=5cm]{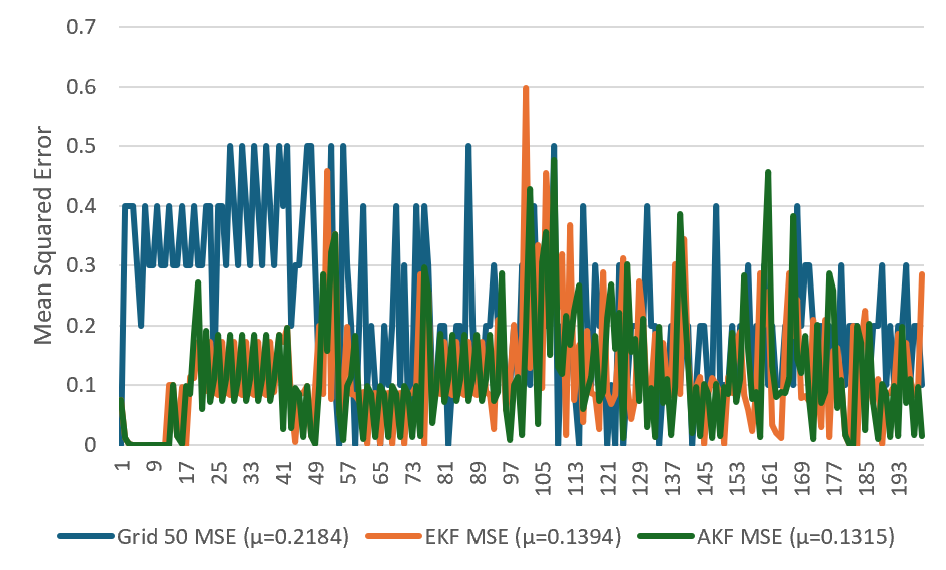} 
\caption{EKF, AKF \& Grid Mean Squared Error}
\end{figure}

The unfiltered Grid algorithm predictions have a modal value $\mu_{Grid}=0.5$ which is close to the cumulative mean $\nu_{cpu}=0.7191$, hence extreme measurement values cause high MSE, an effect more significant at the beginning of the benchmark experiment where CPU utilization is low. This type of error, along with the high variance input data leads to a high cumulative MSE compared to EKF and AKF algorithms. \\

In the Grid algorithm experiments, the benefits of pre-filtering are demonstrated by the significant reduction in cumulative MSE for the filtered Grid algorithms, with the trend $\nu_{Grid} > \nu_{EKFG} > \nu_{AKFG}$.  The exception is a small increase in MSE for Savgol $\nu_{Savgol} > \nu_{Grid}$ which is a much smaller increase when compared to the MSE reduction due to EKFG and AKFG filter algorithms. \\

\begin{figure}[h]
\includegraphics[trim=0 0 30 30,clip,width=0.9\linewidth, height=5cm]{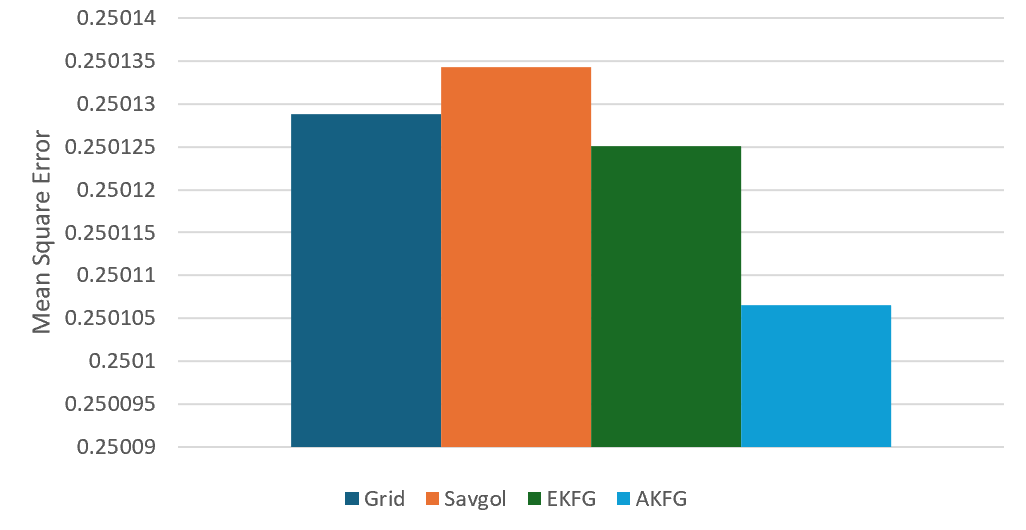} 
\caption{EKFG, AKFG \& Grid All Experiments cumulative MSE}
\end{figure}

The Savgol algorithm does not have a feedback mechanism, hence it does not improve as more data is encountered. The EKF and AKF have an intrinsic advantage in this respect as a data-driven state update executed at every data point. The AKF adds an attention layer to the EKF enabling higher accuracy by identifying distant dependencies in inputs to minimize the impact of recent outliers in the input window. It is demonstrated that the AKF is the most accurate for a small prediction window $\epsilon_{AKF} > \epsilon_{EKF} > \epsilon_{EKF_G} > \epsilon_{Savgol} > \epsilon_{Grid} $.

\begin{table}[h]
\centering
\begin{tabular}{ |p{1.5cm}|p{1.0cm}|p{1.0cm}|p{1.0cm}|p{1.0cm}|p{1.0cm}|  }
 \hline
 \multicolumn{6}{|c|}{ Accuracy } \\
 \hline
 $\epsilon_{Grid}$  & $\epsilon_{Savgol}$ & $\epsilon_{EKF_G}$ & $\epsilon_{AKF_G}$ & $\epsilon_{EKF}$ & $\epsilon_{AKF}$ \\
 \hline
0.74987 & 0.74987 & 0.74988 & 0.74989 & 0.8071 & \textbf{0.8117} \\
 \hline
\end{tabular}
\caption{Google Cloud Charlie Trace Prediction Summary}
 \end{table}

\subsection{Apache Kafka Resource Estimation \& Scaling Evaluation}
Estimating resources for distributed services with API endpoints serving user traffic is difficult, and past resource state can also be unreliable \cite{30}. Hence the resource estimation and auto-scaling impact of AKF-PCA is evaluated using a real-time event processing system implemented on Kafka, a distributed messaging stream processor. Apache Kafka version 2.12-3.3.1 is used in cluster mode with Apache Zookeeper, Kafka broker, and Kafka client nodes \cite{34}.\\

\subsubsection{Kafka Workloads}
In general, requests to web applications including real-time messaging systems have intervals that follow an exponential distribution over small time scales, such as milliseconds. Since they are independent of each other, they can be modeled accurately as a Poisson process \cite{35}. In the first evaluation, a Poisson workload is used to simulate web traffic on a microservice application to perform resource estimation and scaling decisions under real-world conditions. In the subsequent evaluation, multi-day tweet counts from hashtag \#Ukraine at a 1-minute granularity are used to provide a Twitter workload for the microservice application. \\

\subsubsection{Kafka Workload Generation}
Kafka producer nodes are used to generate a workload for the leader broker. In the Poisson workload evaluation, the workload targets the Kafka service endpoint using a single topic with $n$ partitions, where $n = num\_brokers$, the number of brokers. In the Twitter workload evaluation, a real-world Twitter trace is used to create a workload for the Kafka service.  \\

\subsubsection{Attention Layer Training}
The attention layer is trained for both the Poisson and Twitter workloads. The same training scheme is used for both workloads. For the Twitter workload evaluation, the transformer-based attention network is trained using a tweet count prediction task. Tweet counts are collected at a granularity of 1 minute for 5 days for a chosen hashtag. Two training runs were executed at 200 epochs, one for a 2-day training batch, and another for a 3-day training batch. \\




\begin{figure}[h]
\begin{tikzpicture}
\node[inner sep=0pt] (grid_all) at (0,0)
    {\includegraphics[width=.25\textwidth]{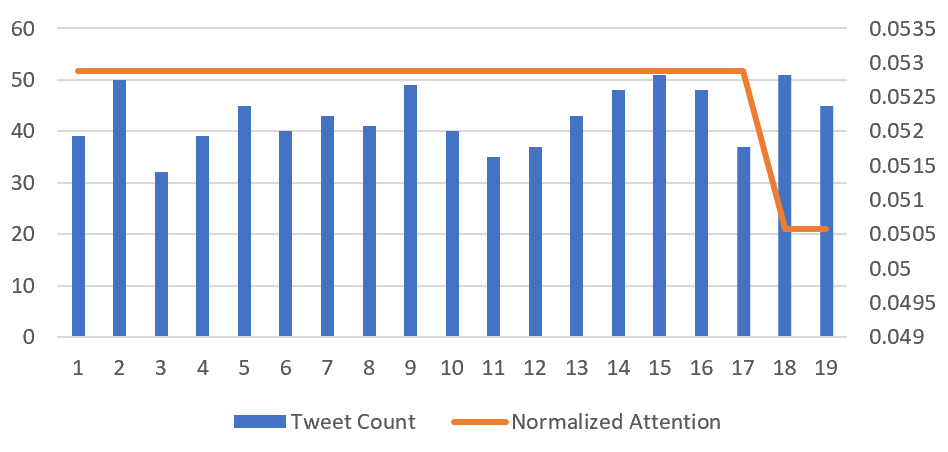}};
\node[inner sep=0pt] (grid_all_t1) at (4.3,0)
    {\includegraphics[width=.25\textwidth]{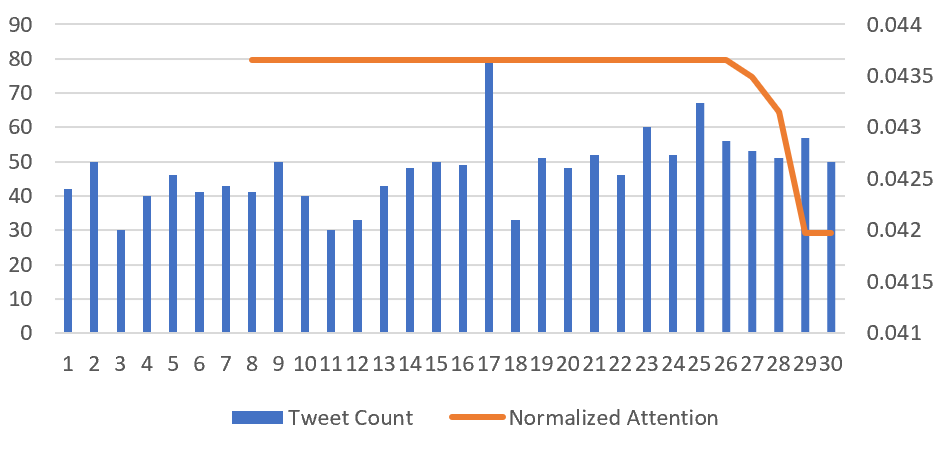}};
\node[inner sep=0pt] (grid_all_t1) at (0,-3.0)
    {\includegraphics[width=.25\textwidth]{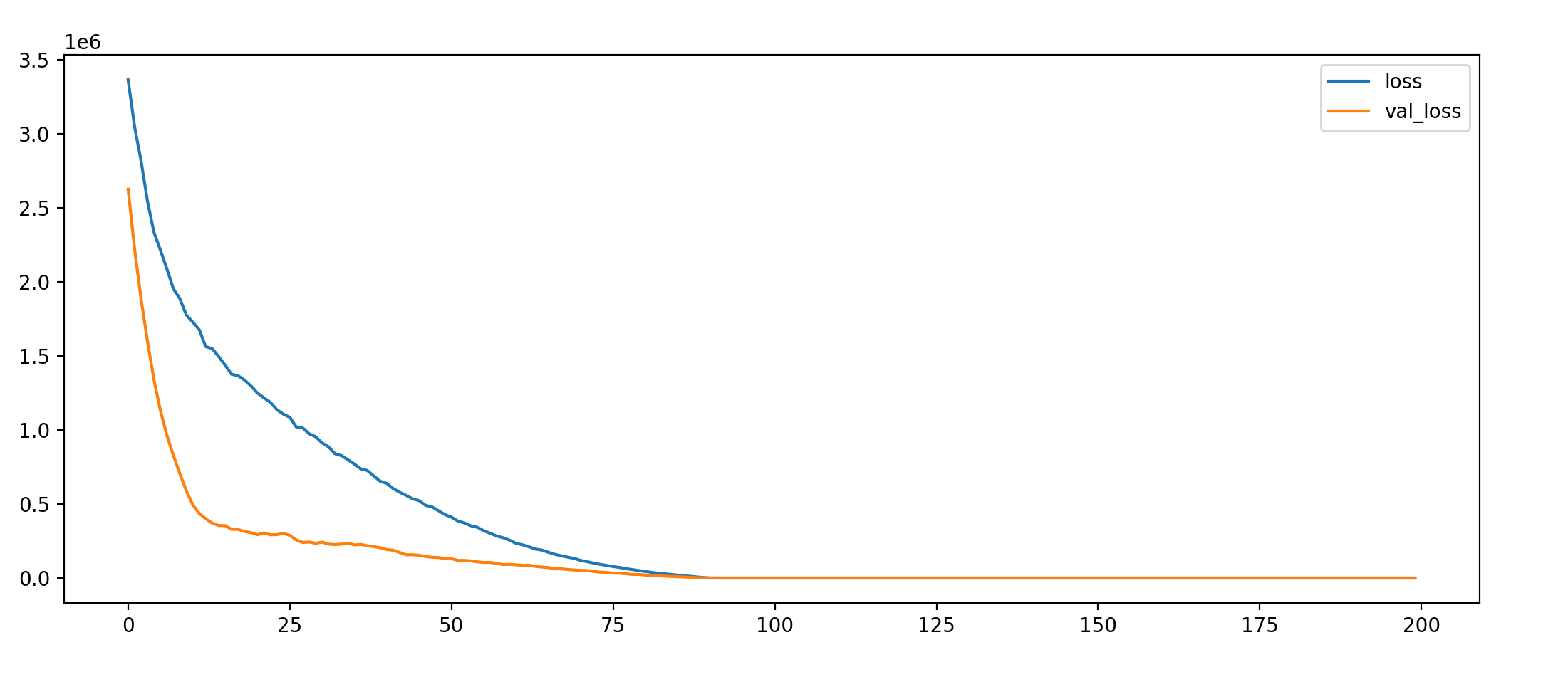}};
\end{tikzpicture}
\caption{2-batch \& 3-batch Tweet count and attention by hashtag in May-2023}
\end{figure}

The transformer network loss function for the 2-day and 3-day batches show that the training performance is very similar with the attention distribution being less evenly distributed in the 3-batch scenario, with the standard deviation $\rho_{2-batch}=0.00725$ and $\rho_{3-batch}=0.000487$ $\implies \rho_{2-batch} \gg \rho_{3-batch}$ demonstrating a less uniform attention weight distribution in the 3 batch scenario. This effectively increases input filtering when more training data is provided. 
\\

\subsubsection{Studying the Twitter Dataset with a Tweet Count Prediction Task}
To test the efficacy of the attention training method, the accuracy of UKF, EKF-PCA, and AKF-PCA are evaluated in the task of predicting tweet counts. With a granularity of 1 minute, the history of tweet counts and timestamps over a 3-day window is used to predict a future 120-second window. The error $\epsilon$ is shown in the prediction statistics, where for a set of $n$ tweets $t_{i=1\to n}$ and predictions $\hat{t}_{i=1\to n}$

\[
\epsilon = \frac{1}{n}\sum_{i}^{n}\frac{|\hat{t}_{i}-t_{i}|}{t_{i}}
\]

\begin{table}[h]
\centering
\begin{tabular}{ |p{1.5cm}|p{1.5cm}|p{1.5cm}|  }
 \hline
 \multicolumn{3}{|c|}{Prediction Error } \\
 \hline
 $\epsilon_{UKF}$  & $\epsilon_{EKF-PCA}$ & $\epsilon_{AKF-PCA}$ \\
 \hline
\textbf{0.1835} & 0.1862 & 0.1857 \\
 \hline
\end{tabular}
\caption{Prediction statistics for the Twitter Workload}
 \label{tbl:tbl1}
 \end{table}

The relation $\epsilon_{UKF}$ $<$ $\epsilon_{AKF-PCA}$ $<$ $\epsilon_{EKF-PCA}$ is observed, which shows the improvement of AKF-PCA over EKF-PCA regarding prediction error, while also showing that in this task, UKF has the lowest error. The application of attention to UKF for this task is worth exploring in the future.

\subsubsection{Kafka Topic \& Partitions}
There is one topic, \textit{topic1}, and a configured initial set of $n$ partitions. As scaling actions are performed by the autoscaler, the $n$ increases as new brokers are added. Each consumer is subscribed to a single partition, and a specialized estimator node queries \textit{Zookeeper} to determine the leader broker identity when a scaling operation is executed. 

\begin{figure}[h]
\includegraphics[width=0.9\linewidth, height=5cm]{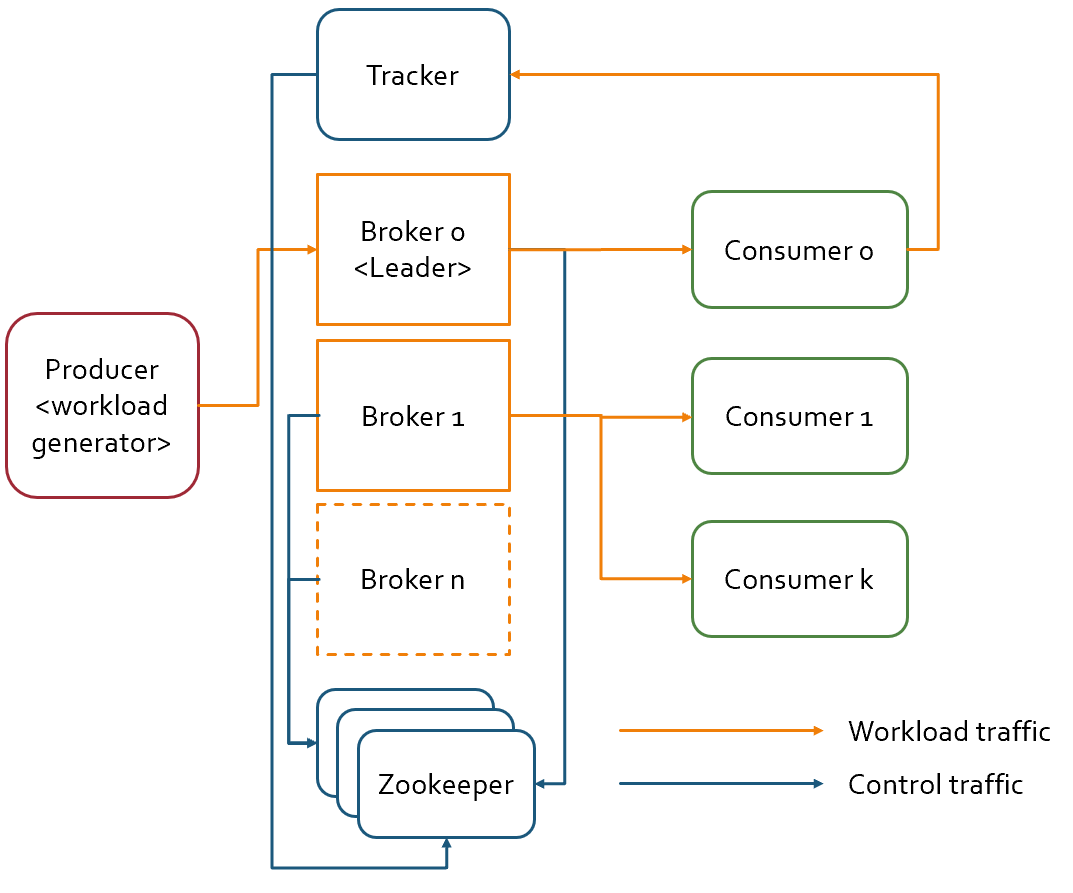}
\caption{Kafka Estimator Experiment Setup}

\label{fig:fig6}
\end{figure}

\subsubsection{Kafka Estimator Node}
Kafka consumers are subscribed to  \textit{topic1}, with a single consumer dedicated to updating the \textit{estimator} node. This consumer updates the \textit{estimator} node on every notification. The estimator node predicts consumer latency and throughput for all consumers at each iteration. \\

At the configured update rate $r = 0.25$ the estimator provides measurement updates to an AKF-PCA thread using the Kalman filter update algorithm. This means one of every $\frac{1}{r}$ iterations will cause an update. The tracker consumer node is subscribed to all partitions to estimate all brokers' metrics. \\

\subsubsection{Kafka Consumers}
The rest of the consumers only do periodic updates at the rate $r$ based on the received message rate The \textit{estimator} then estimates the consumer latency of all partitions from these periodic updates using AKF-PCA. \\

\subsubsection{Kafka Latency Estimation}

Using these latency estimates, the \textit{estimator} performs threshold-based scaling. When the estimated latency passes the configured threshold, a new broker is added to the Kafka cluster. This only happens once during the experiment duration. \\

In the results, the producer-to-estimator AKF-PCA-measured local latency has a uniform distribution with a mean frequency of 962.93 requests every 0.4ns, and a narrow maximum-minimum frequency variance of 790 requests every 0.4ns. The estimated global latency has a less uniform distribution with a mean frequency of 1022.72 requests every 0.4ns, and a wide maximum-minimum frequency variance of 22500 requests every 0.4ns, due to a 22 times reduction in latency variance after the estimator provisioning action which involves launching a new broker node. This indicates latency variance reduction after the provisioning action. \\

In addition, a comparison of the measured local latency to the estimator-estimated global latency is done using the \textit{residual variance} measure. This involves computing a \textit{least squares linear regression} and associated variance

\begin{equation}
\rho_{t} = \mathbb{E}[(y_{t}-\hat{y_{t}})^{2}]  
\end{equation} 

$\rho_{t}$ refers to the residual variance of signal type $t$ and $\hat{y_{t}}$ is the linear regression solution of $y_{t}$, the latency time series of signal type $t$. The estimator-measured latency residual variance $\rho_{z} = 1.716 \mu s^{2}$ is significantly higher than the estimated latency variance $\rho_{x} = 0.04966 \mu s^{2}$, showing a $97\%$ noise reduction. \\



\begin{figure}[h]
\begin{tikzpicture}
\node[inner sep=0pt] (grid_all) at (0,0)
    {\includegraphics[width=.25\textwidth]{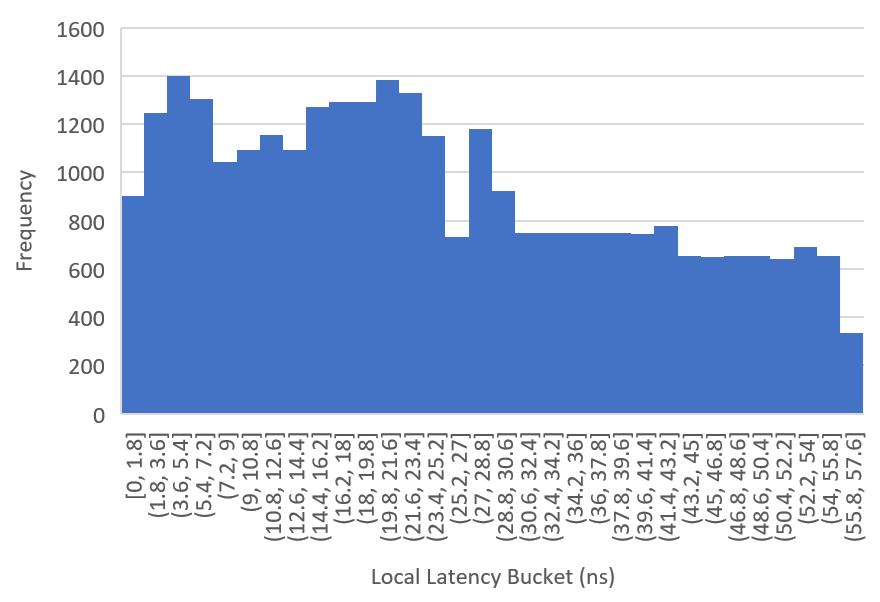}};
\node[inner sep=0pt] (grid_all_t1) at (4.3,0)
    {\includegraphics[width=.25\textwidth]{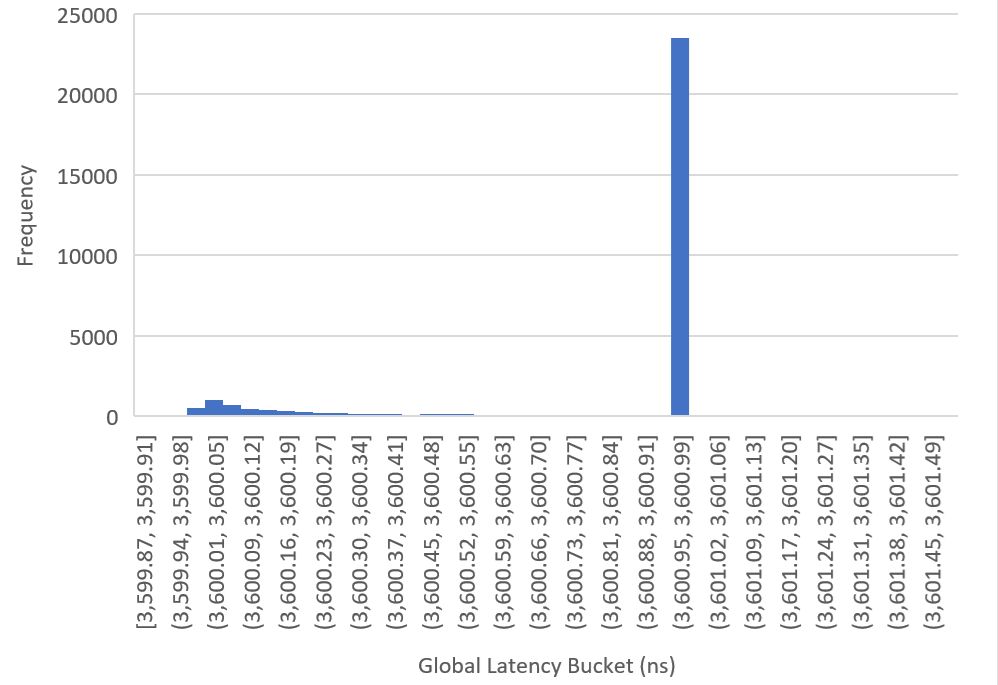}};
\end{tikzpicture}
\caption{AKF-PCA Measured Local \& Estimated Global Latency Distribution}
\end{figure}

\begin{figure}[h]
\begin{tikzpicture}
\node[inner sep=0pt] (thresholds) at (0,0)
    {\includegraphics[width=0.50\textwidth]{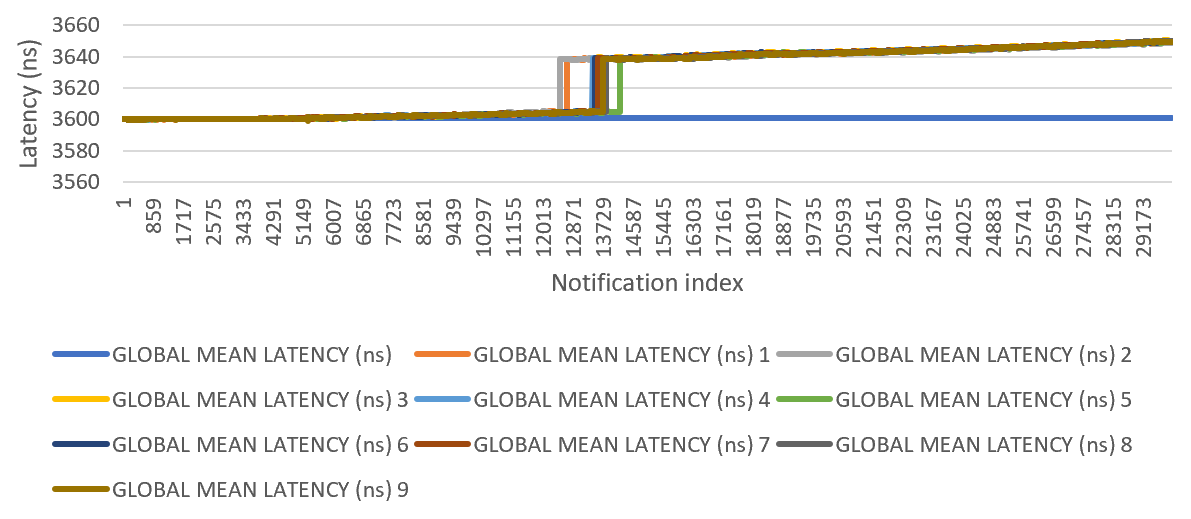}};
\draw[draw=red] (-0.35,0.9) rectangle ++(0.6,1.0);
\draw[<->,thick,draw=red] (-0.39,0.75) -- (0.33,0.75) node[midway,fill=white,text=black] {$\sigma$};;
\end{tikzpicture}
\caption{AKF-PCA Multi-iteration Threshold Estimation}
\end{figure}

\subsubsection{Kafka Latency Threshold Detection}
The stability of the scaling action in response to noisy traffic data is an important aspect of the quality of an automated scaling system. The scaling operation of the estimator is examined repeatedly over multiple iterations, with the number of iterations $n_{iter} = 10$. A single scaling action is permitted in each iteration. \\

The scaling action triggered by message latency crossing the configured threshold requires a duration $d_{s} \approx 40 \mu s$ which causes messages to queue on the leader broker. Once the scaling operation is done the messages resume delivery across both partitions, causing a step increase in latency of each message of $d_{s}$ delivered after scaling. \\

\subsubsection{Kafka Experiments Discussion}
The scaling action time is represented by a step increase in the message latency plot. The variance of the scaling action time is an indicator of detector quality, with a narrow variance showing more stable detection regarding the number of requests. The variance is computed as $  \sigma_{i} = \mathbb{E}[(t_{i}-\Bar{t_{i}})^{2}] $, where $t_{i}$ refers to the initiation time of the scaling action, $\Bar{t_{i}}$ is the mean of $t_{i}$ over all iterations, and $i$ refers to the experiment type (AKF-PCA, EKF-PCA, UKF, Passive), computed for a set of experiments involving the different estimators. \\

The \textit{passive} estimator refers to a simple threshold-crossing detector using the raw measured message latency to make scaling decisions. The AKF-PCA estimator, with the lowest variance in scaling action time, is the highest quality estimator in this task.
 
\begin{table}[t]
\centering
\begin{tabular}{ |p{1.5cm}|p{1.5cm}|p{1.5cm}|p{1.5cm}|  }
 \hline
 \multicolumn{4}{|c|}{Threshold Variance ($\mu s^{2}$)} \\
 \hline
 $\sigma_{PASSIVE}$ & $\sigma_{UKF}$  & $\sigma_{EKF-PCA}$ & $\sigma_{AKF-PCA}$ \\
 \hline
1042.0 & 700.5 & 294.0 & \textbf{17.25} \\
 \hline
\end{tabular}
\caption{Kafka scaling statistics using the Poisson Workload}
 \label{tbl:tbl3}
 \end{table}
 
$\implies \sigma_{AKF-PCA} \ll \sigma_{EKF-PCA} \ll \sigma_{UKF} \ll \sigma_{PASSIVE}$. \\

The same trend is seen in the Twitter workload experiment. As the workloads are generated at a controlled rate, the threshold detection latency can be used to explicitly calculate the total number of notifications sent by the workload generator in the Kafka producer to the broker nodes. Hence the threshold detection stability is a direct indicator of the workload-based auto-scaler stability.

\begin{table}[t]
\centering
\begin{tabular}{ |p{1.5cm}|p{1.5cm}|p{1.5cm}|p{1.5cm}|  }
 \hline
 \multicolumn{4}{|c|}{Threshold Variance ($\mu s^{2}$)} \\
 \hline
 $\sigma_{PASSIVE}$ & $\sigma_{UKF}$  & $\sigma_{EKF-PCA}$ & $\sigma_{AKF-PCA}$ \\
 \hline
10669.3 & 793.0 & 391.5 & \textbf{264.25} \\
 \hline
\end{tabular}
\caption{Kafka scaling statistics using the Twitter Workload}
 \label{tbl:tbl4}
 \end{table}

\begin{figure}[h]
\includegraphics[clip,width=1\linewidth]{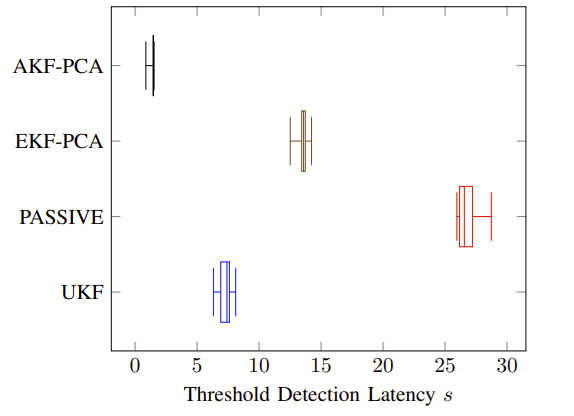}
\caption{Kafka detection scaling stability - Poisson Workload}
\label{fig:poisson}
\end{figure}

\begin{figure}[h]
\includegraphics[clip,width=1\linewidth]{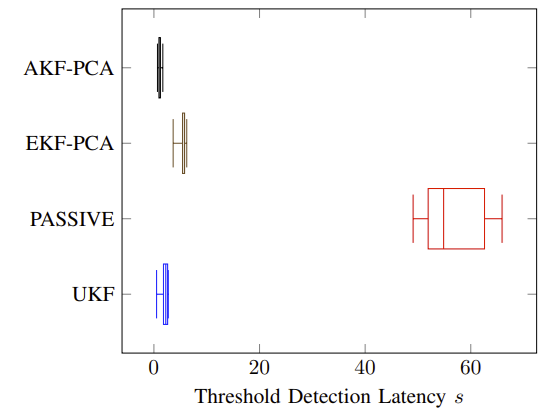}
\caption{Kafka detection scaling stability - Twitter Workload}
\label{fig:twitter}
\end{figure}

\pagebreak
\newpage

\section{Conclusions}

In this work, we present an algorithm to solve the problem of low-latency prediction using time series data with noise and extreme points, in the form of an improvement to the Extended Kalman filter using a novel AKF-PCA Kalman filter approach with PCA and \textit{attention}. We explore a joint UKF-PCA and EKF-PCA method. AKF-PCA uses a transformer network to train an attention layer which is combined with the Kalman filter to improve estimation accuracy. This has been applied to an Apache Kafka controller and demonstrated significant improvement in the workload-based auto-scaler stability.  \\

For resource estimation tasks, the standalone EKF-PCA and UKF-PCA approaches show improved accuracy over the EKF and UKF primarily in cases where measurement data has outliers and poorer performance for the higher variance simulated time series. The \textit{$\epsilon$ correction} heuristic is developed for estimator selection using signal variance bounds. \\

The Google Cloud benchmark experiments show the performance impact of Kalman filters in noise and extreme point mitigation through pre-filtering for the Bidirectional Grid LSTM neural network model, and in short-window prediction tasks. The EKF has better performance as an input filter for BG-LSTM while the AKF has the best resource prediction accuracy: $\epsilon_{AKF} > \epsilon_{EKF} > \epsilon_{EKF_G} > \epsilon_{Savgol} > \epsilon_{Grid} $. The improvements stem from the feedback mechanism present in the Kalman filter models during inference, the suitability of the KF algorithm to high variability data, and the effectiveness of attention in mitigating extreme points. \\ 

The Kalman filter is an effective model estimation method using noisy time series data. We demonstrate applications in resource estimation for real-time message processing using Kafka. Improved resource estimation for Kafka using AKF-PCA when compared to EKF-PCA, UKF, and passive threshold detection has been demonstrated in terms of both noise reduction and better workload-based stability with $\sigma_{AKF-PCA}$ to $\sigma_{UKF}$  reduction of $66 \%$ on the Twitter workload. Concerning the number of requests processed when scaling occurs, the auto-scaler has improved stability with AKF-PCA compared to EKF, EKF-PCA, UKF, and passive threshold detection.  \\

Improved auto-scaler stability guarantees stricter and more reliable Service Level Agreements (SLA) \cite{1}. Low stability can result not only in violation of SLA but also in over-provisioning of resources and increased scaling costs. Request volume-based SLAs are particularly affected by auto-scaler stability, and the attention-based computational complexity of AKF-PCA is justifiable in this regard when compared to the state-of-the-art. \\

Future work involves using neural networks to estimate parameters for incomplete Kalman filter models. A new Kafka estimator-based performance evaluation for large cluster systems of varying topology and multiple filters will be examined. It is also of interest to explore the bounds of the Kafka estimator hyperparameters necessary for optimally stable auto-scalers and to explore the potential performance gains brought to bear by Kalman filter variability and noise mitigation on state-of-the-art multi-modal neural network models.  \\


\end{document}